\documentclass{article} 
\usepackage{apjfonts,epsf,pstricks,natbib}
\usepackage[onecolumn]{emulateapj}
\bibpunct[;]{(}{)}{;}{a}{ }{,}

\newcommand{\bg}[1]{\mbox{\boldmath $#1$}}
\newcommand{\me}{m_{\rm e}}
\newcommand{\te}{T_{\rm e}}
\newcommand{\mus}{\mu_{\rm s}}

\begin{document}

\submitted{To appear in the v543n2 Nov 10, 2000 issue of the
Astrophysical Journal} 

\righthead{SAZONOV \& SUNYAEV}
\lefthead{NARROW LINE AFTER SINGLE SCATTERING}

\title{THE PROFILE OF A NARROW LINE AFTER SINGLE SCATTERING BY
MAXWELLIAN ELECTRONS:\\ RELATIVISTIC CORRECTIONS TO THE KERNEL OF THE
INTEGRAL KINETIC EQUATION}

\author{Sergei Y. Sazonov and Rashid A. Sunyaev}

\affil{Max-Planck-Institut f\"ur Astrophysik,
Karl-Schwarzschild-Str.~1, 86740 Garching bei M\"unchen, Germany}

\affil{Space Research Institute (IKI), Profsoyuznaya~84/32, Moscow
117810, Russia} 

\begin{abstract}
The photon frequency distribution that results from single
Compton scattering of monochromatic radiation on thermal electrons is derived
in the mildly relativistic limit. Algebraic expressions are given for (1)
the photon redistribution function, $K(\nu,\bg{\Omega}\rightarrow
\nu^\prime,\bg{\Omega^\prime})$, and (2) the spectrum produced in the
case of isotropic incident radiation, $P(\nu\rightarrow\nu^\prime)$. The former
is a good approximation for electron temperatures 
$k\te\lesssim 25$~keV and photon energies $h\nu\lesssim 50$~keV, and the
latter is applicable when $h\nu(h\nu/\me c^2)\lesssim k\te\lesssim 25$~keV, 
$h\nu\lesssim 50$~keV. Both formulae can be used for describing the profiles
of X-ray and low-frequency lines upon scattering in hot, optically thin
plasmas, such as present in clusters of galaxies, in the coronae of
accretion disks in X-ray binaries and active galactic neclei (AGNs),
during supernova explosions, etc. Both formulae can also be employed
as the kernels of the corresponding 
integral kinetic equations (direction-dependent and isotropic) in the
general problem of Comptonization on thermal electrons. The
$K(\nu,\bg{\Omega}\rightarrow \nu^\prime,\bg{\Omega^\prime})$ kernel, in
particular, is applicable to the problem of induced Compton interaction of
anisotropic low-frequency radiation of high brightness temperature with free
electrons in the vicinity of powerful radiosources and
masers. Fokker-Planck-type expansion (up to fourth order) of the
integral kinetic equation with the $P(\nu\rightarrow\nu^\prime)$
kernel derived here leads to a generalization of the Kompaneets
equation. We also present (1) a simpler kernel that is necessary and
sufficient to derive the Kompaneets equation and (2) an expression
for the angular function for Compton scattering in a hot plasma, which includes
temperature and photon energy corrections to the Rayleigh angular function.
\end{abstract}


\section{INTRODUCTION}

A monochromatic spectral line will be broadened after single Compton
scattering on thermal electrons in an optically thin, hot plasma. The
emergent spectrum will depend on the angle between the direction,
$\bg{\Omega}$, from which the photons are supplied and the observer's
direction, $\bg{\Omega^\prime}$. A classical problem then arises of
finding the redistribution function, $K(\nu,\bg{\Omega}\rightarrow
\nu^\prime,\bg{\Omega^\prime})$, which gives the probability for the
incident photon $(\nu,\bg{\Omega})$ to be scattered in the direction
$\bg{\Omega^\prime}$ with a frequency of $\nu^\prime$ \citep{dirac25}.

In the case in which the incident radiation is isotropic and monochromatic, the
spectrum resulting from single Compton scattering can be found by
integrating $K(\nu,\bg{\Omega}\rightarrow \nu^\prime,\bg{\Omega^\prime})$
over the scattering angle ($\mus=\bg{\Omega}\bg{\Omega^\prime}$):
\begin{equation}
P(\nu\rightarrow\nu^\prime)=\int K(\nu,\bg{\Omega}\rightarrow
\nu^\prime,\bg{\Omega^\prime})\,d\bg{\Omega^\prime}=
2\pi\int K(\nu,\bg{\Omega}\rightarrow
\nu^\prime,\bg{\Omega^\prime})\,d\mus .
\label{k_int}
\end{equation}

Spectra described by equation (\ref{k_int}) may in particular form when
spherical symmetry is present in the system. Consider, for example, an
isotropic source of monochromatic radiation that is either located
at the center of a spherical cluster of galaxies with hot intergalactic gas or 
distributed spherically symmetrically across the cluster. The frequency
distribution of photons that have experienced a single scattering
within the cluster will then be precisely
$P(\nu\rightarrow\nu^\prime)$ if the entire cluster is probed at 
the same time. If instead the spectrum is collected from a part of the
cluster, then one must take into account the angular distribution of
incident photons and make use of the direction-dependent
$K(\nu,\bg{\Omega}\rightarrow \nu^\prime,\bg{\Omega^\prime})$ function.

Single-scattering line profiles, which can provide important
information on the conditions (particularly the temperature) of the plasma
in the source, can originate in various astrophysical environments. For the
X-ray spectral band, these include intracluster gas, the coronae of
accretion disks around black holes in binary stellar systems and
active galactic nuclei (AGNs), and plasma streams outflowing from the
neutron star during super-Eddington X-ray bursts in bursters.

High-quality spectroscopic X-ray observations necessary for the
detection and measurement of such lines will soon become possible
with the satellites {\sl Chandra}, {\sl XMM} (both already in orbit),
and {\sl Spectrum-X-Gamma}, and in a more distant outlook, with
{\sl Constellation-X} and {\sl XEUS}. This was one of the primary
motives for us in initiating the current study, in which we calculate
analytically the functions $K(\nu,\bg{\Omega}\rightarrow
\nu^\prime,\bg{\Omega^\prime})$ (eq. [\ref{k_set}]) and
$P(\nu\rightarrow\nu^\prime)$ (eqs. [\ref{p_set}] and [\ref{p_rn}]) in the
mildly relativistic limit.

\subsection{Integral Kinetic Equation}

Apart from the single-scattering problem outlined above,
$K(\nu,\bg{\Omega}\rightarrow \nu^\prime,\bg{\Omega^\prime})$ 
and $P(\nu\rightarrow\nu^\prime)$ can be used as the kernels of the
corresponding integrodifferential kinetic equations describing the
Comptonization of photons on Maxwellian electrons. In the anisotropic
problem, the kinetic equation in the case of an infinite homogeneous medium
can be written as
\begin{eqnarray} 
\frac{\partial I(\nu,\bg{\Omega},\tau)}{\partial\tau}=-\int
I(\nu,\bg{\Omega},\tau)
K(\nu,\bg{\Omega}\rightarrow\nu^\prime,\bg{\Omega^\prime}) 
[1+n(\nu^\prime,\bg{\Omega^\prime},\tau)] d\nu^{\prime} d\bg{\Omega^\prime}
\nonumber\\
+\int \frac{\nu}{\nu^\prime}I(\nu^\prime,\bg{\Omega^\prime},\tau)
K(\nu^\prime,\bg{\Omega^\prime}\rightarrow\nu,\bg{\Omega})
[1+n(\nu,\bg{\Omega},\tau)] d\nu^\prime d\bg{\Omega^\prime},
\label{kinetic_k}
\end{eqnarray}
where $I(\nu,\bg{\Omega})$ is the specific intensity of the radiation,
$n=c^2 I/(2h\nu^3)$ is the occupation number in photon phase space,
and $\tau=\sigma_{\rm T} N_{\rm e} ct$ is the dimensionless time
($\sigma_{\rm T}$ is the Thomson scattering cross section and $N_{\rm
e}$ is the number density of electrons). It is easy to write down kinetic
equations similar to equation (\ref{kinetic_k}) for any of the
standard problems of radiation transfer. 

In the isotropic case, the kinetic equation becomes
\begin{equation}
\frac{\partial I(\nu,\tau)}{\partial\tau}=-\int I(\nu,\tau) P(\nu
\rightarrow \nu^\prime)[1+n(\nu^\prime,\tau)] d\nu^{\prime}+
\int\frac{\nu}{\nu^\prime} I(\nu^\prime,\tau) P(\nu^\prime \rightarrow
\nu)[1+n(\nu,\tau)] d\nu^\prime.
\label{kinetic}
\end{equation}

The integrodifferential equation (\ref{kinetic}) can in general be solved
numerically if its kernel, $P(\nu\rightarrow\nu^\prime)$, is
known. Alternatively, one can treat Comptonization problems using
Monte Carlo methods \citep[see, e.g., the review by][]{pozetal83}. In the
nonrelativistic limit, i.e. when the typical photon energy, $h\nu$,
and the plasma temperature, $k\te$, are both negligibly small
compared to the electron rest energy, $\me c^2$, the variation in
intensity at a given frequency is largely governed by transitions in
a narrow interval of a continuum spectrum near this frequency. If the initial
distribution of the photons in frequency is smooth enough (in a problem with a
small number of scatterings) or the formation of a spectrum by multiple
scatterings is studied, it is possible to perform a Fokker-Planck-type
expansion of the integral equation (\ref{kinetic}), thereby reducing it to a
much simpler differential equation that describes the diffusion and
flow of the photons in frequency space:
\begin{eqnarray}
\frac{\partial n(\nu)}{\partial \tau}=\frac{1}{\nu^2}\frac{\partial}
{\partial\nu}\left\{-\nu^2n\langle\Delta\nu\rangle(1+n)+
\frac{1}{2}\left[-\nu^2n\langle (\Delta\nu)^2\rangle\frac{\partial n}
{\partial\nu}+(1+n)\frac{\partial}{\partial\nu}
\nu^2n\langle(\Delta\nu)^2\rangle\right]
\right.
\nonumber\\
\left.
+\frac{1}{6}\left[-\nu^2n\langle(\Delta\nu)^3\rangle
\frac{\partial^2 n}{\partial\nu^2}+\frac{\partial n}{\partial\nu}
\frac{\partial}{\partial\nu}\nu^2n\langle(\Delta
\nu)^3\rangle-(1+n)\frac{\partial^2}{\partial\nu^2}
\nu^2n\langle(\Delta\nu)^3\rangle\right]
\right.
\nonumber\\
\left.
+\frac{1}{24}\left[-\nu^2n\langle (\Delta\nu)^4\rangle\frac{\partial^3n}
{\partial\nu^3}+\frac{\partial^2n}{\partial\nu^2}
\frac{\partial}{\partial\nu}\nu^2n\langle(\Delta\nu)^4\rangle
-\frac{\partial n}{\partial\nu}\frac{\partial^2}{\partial\nu^2}
\nu^2n\langle(\Delta\nu)^4\rangle
\right.\right.
\nonumber\\
\left.\left.
+(1+n)\frac{\partial^3}{\partial\nu^3}\nu^2n\langle(\Delta\nu)^4\rangle
\right]+...\right\}.
\label{fokker}
\end{eqnarray}

The moments of the kernel that enter this equation are found from
the formula
\begin{equation}
\langle(\Delta\nu)^n\rangle=\int P(\nu\rightarrow\nu^\prime)
(\nu^\prime-\nu)^n d\nu^\prime.
\label{mom1}
\end{equation}
Substituting the first two moments, $\langle\Delta\nu\rangle$ and
$\langle(\Delta\nu)^2\rangle$, calculated to an accuracy of
$k\te/\me c^2$ and $h\nu/\me c^2$ (using the kernel given by eq. [\ref{p_k}]
below) into the corresponding terms of equation (\ref{fokker}) leads to the
famous \cite{kompaneets57} equation:
\begin{eqnarray}
\frac{\partial n(\nu)}{\partial\tau}=\frac{h}{\me c^2}\frac{1}{\nu^2}
\frac{\partial}{\partial \nu}\nu^4\left(n+n^2+\frac{k\te}{h}
\frac{\partial n}{\partial\nu}
\right).
\label{komp}
\end{eqnarray} 
The Kompaneets equation is valid in the nonrelativistic limit ($h\nu$,
$k\te\ll \me c^2$). The last parenthesized term in equation (\ref{komp})
describes the frequency diffusion of photons due to the Doppler
effect and the transfer of energy from the electrons to the radiation; the
first term describes the downward photon flow along the frequency axis due
to Compton recoil, and the second term, which also is owing to recoil,
accounts for induced Compton scattering.

In studying the interaction between energetic photons and hot electrons
($h\nu,\,\,k\te\gtrsim 0.01 \me c^2$), relativistic corrections to the
kernel and the Kompaneets equation become important. In order to find the
main corrections (of the order of $h\nu/\me c^2$ and $k\te/\me c^2$),
it proves necessary to take into account in equation (\ref{fokker}) all terms
that depend on the first four moments of the kernel. The resultant
generalization of the Kompaneets equation was found by \cite{itoetal98} and
\cite{chalas98} when these authors examined the distortion of the spectrum of
the cosmic microwave background (CMB) during interaction with the hot
intergalactic gas in clusters of galaxies. \cite{rephaeli95} pointed out the
important role of corrections of the order of $k\te/\me c^2$ in this
phenomenon. The kernel $P(\nu\rightarrow \nu^\prime)$ was not given in
explicit form in the derivation of \cite{itoetal98} and \cite{chalas98}.
Instead, the Fokker-Planck operator was applied to the kinetic equation
written in the most general form, where the amplitude appears for a
transition from the initial state with given 4-momenta of the photon and
electron into the final state with the corresponding 4-momenta. Earlier,
\cite{rosetal78} and \cite{illetal79} added to the Kompaneets equation a
dispersion term associated with Compton recoil and a term that takes into
account, in a first approximation, the transition from the Thomson
cross section to the Klein-Nishina one. Their equation, which describes much
better than the Kompaneets equation the scattering of energetic photons
($h\nu\sim 0.1\me c^2$) by sufficiently cold electrons ($k\te\ll h\nu$), is
a particular case of the more general formula of \cite{itoetal98} and
\cite{chalas98}.

Despite the attractiveness of employing the Fokker-Planck
approximation in treating Comptonization problems, its scope is limited. 
For example, in considering the effect of Comptonization from a small
number of scatterings on the profiles of narrow spectral lines, the initial
radiation spectrum cannot be represented as a finite Taylor series in terms
of the frequency variation, and, therefore, Fokker-Planck-type equations are
not applicable. For this kind of problem, it is necessary to make use of the
integral kinetic equations (\ref{kinetic_k}) or (\ref{kinetic}), which
requires knowledge of their kernels, $K(\nu,\bg{\Omega}\rightarrow
\nu^\prime,\bg{\Omega^\prime})$ and $P(\nu\rightarrow\nu^\prime)$. 

\subsection{The Kernel}

\cite{dirac25} has given an approximate expression for the direction-dependent
kernel, $K(\nu,\bg{\Omega}\rightarrow \nu^\prime,\bg{\Omega^\prime})$. The
Doppler shift was taken into account to within the first order in
$v/c$, but Compton recoil was totally neglected. Integration of Dirac's
$K(\nu,\bg{\Omega}\rightarrow \nu^\prime,\bg{\Omega^\prime})$ function over
the scattering angle results in the zero-order approximation for the
isotropic kernel $P(\nu\rightarrow\nu^\prime)$ \citep[][see also Weymann
1970]{hummih67}. Being symmetric in frequency variation [$P(\nu\rightarrow
\nu+\Delta\nu)=P(\nu\rightarrow \nu-\Delta\nu)$], this kernel does not take
into account the average Doppler increment in the photon energy: $\langle
\Delta\nu/\nu\rangle=4k\te/\me c^2$. With the zero-order approximation for
the kernel, it is not possible to describe many important astrophysical
phenomena, such as:

(a) distortion of the spectrum of the CMB in the
direction of galaxy clusters \citep[][see the recent review by
Birkinshaw 1999]{sunzel72},

(b) The $y$ \citep{zelsun69} and Bose-Einstein $\mu$ \citep{sunzel70}
distortions of the CMB spectrum resulting from energy release
in the early universe (see the reviews by Danese \& de Zotti 1977; Sunyaev
\& Zeldovich 1980; the book by Peebles 1993; and the strict
constraints on these distortions placed by the Far-Infrared Absolute
Spectrophotometer on board COBE, Fixsen et al. 1996),

(c) the formation of hard power-law tails in the emission spectra of the
famous X-ray source Cygnus X--1 \citep{suntru79} and other stellar-mass
black hole candidates, AGNs, and quasars; and in the
spectra of accreting neutron stars \citep[e.g.,][]
{shaetal76,suntit80,pozetal83,tanshi96,pousve96,naretal98,zdzetal98}.

Babuel-Peyrissac \& Rouvillois (1970, hereafter BR70) have obtained a
more accurate expression for the $K(\nu,\bg{\Omega}\rightarrow
\nu^\prime,\bg{\Omega^\prime})$ kernel, by taking into account not only
the average Doppler and recoil frequency shifts (the latter is $\langle
\Delta\nu/\nu\rangle=-h\nu/\me c^2$), but also, in the first order,
Klein-Nishina corrections to the scattering cross section. BR70 also
attempted to find relativistic corrections due to high electron velocities
($k\te\sim 0.1\me c^2$), but did not include all relevant terms (as we show
in the present paper), and therefore their expression is valid for
nonrelativistic electrons only. The allowance for Compton recoil is also
made in the $K(\nu,\bg{\Omega}\rightarrow \nu^\prime,\bg{\Omega^\prime})$
kernel written down by \cite{zeletal72}. Integration of the BR70 kernel over
the scattering angle ignoring recoil ($h\nu\ll k\te\ll\me c^2$) leads to the
first-order approximation for the $P(\nu\rightarrow\nu^\prime)$ kernel
\citep{sunyaev80}. The formula of \cite{sunyaev80}, as opposed to
the approximation of \cite{hummih67}, accounts for the asymmetry of the
scattered line profile and enables us to describe (in the nonrelativistic
limit) the above-mentioned phenomena related to the transfer of energy from
hot electrons to radiation. Using this expression, one can derive the
diffusion part of the Kompaneets operator (the last term of eq.
[\ref{komp}]). The remaining two (flow) terms can be found by applying the
thermodynamic principle that the desired equation must obey: a Planckian 
distribution of photons must remain unchanged during the interaction with
electrons of the same temperature. This is the method followed by the
authors of the original derivation of the Kompaneets equation for finding
not only the term describing the downward motion of the photons along the
frequency axis, but also the term describing the effect of induced scattering
(report No. 336 of the Chemical Physics Institute of USSR Academy of
Sciences 1950; Kompaneets 1957; Ya.B. Zeldovich 1968, private communication).

\cite{keretal86} have derived a semianalytical relativistic formula for the
$K(\nu,\bg{\Omega}\rightarrow\nu^\prime,\bg{\Omega^\prime})$ kernel that is 
valid for arbitrary values of photon energy and electron temperature.
With this formula, the calculation of the kernel is reduced to computing
a single integral over the electron Lorentz factor. \cite{keretal86} further
succeded in expanding their basic expression in powers of $k\te/\me c^2$, and
thus derived an algebraic formula for
$K(\nu,\bg{\Omega}\rightarrow\nu^\prime,\bg{\Omega^\prime})$ that is 
applicable in the low-temperature limit $k\te\lesssim 0.1\me c^2$, with no  
limitation imposed on the photon energy. 

Other previous relativistic studies of the Compton scattering kernel also
need to be mentioned. \cite{ahaato81} were the first to derive the exact
relativistic formula for the direction-dependent kernel,
$K(\nu,\bg{\Omega}\rightarrow\nu^\prime,\bg{\Omega^\prime})$, for an 
isotropic ensemble of monoenergetic electrons. \cite{nagpou94} 
(see also references therein) derived the exact relativistic expression for
the isotropic kernel, $P(\nu\rightarrow\nu^\prime)$, again for an ensemble
of monoenergetic electrons. These formulae make it possible to 
solve efficiently very general Comptonization problems, i.e., with no
constraints set on the parameters. However, their application always implies
the need to perform one or two numerical integrations, usually over a
specified distribution of electron energies (e.g., Maxwellian), and
sometimes over the scattering angle. We also note that the scattering
kernel has been the subject of a number of studies that employed numerical
methods \citep[e.g.,][]{pomraning73,illetal79,pozetal79,loeetal91,molbir99}.

In the present paper, we have obtained algebraic approximate expressions for
the kernels $K(\nu,\bg{\Omega}\rightarrow\nu^\prime,\bg{\Omega^\prime})$ (eq.
[\ref{k_set}]) and $P(\nu\rightarrow\nu^\prime)$ (eqs.
[\ref{p_set}] and [\ref{p_rn}]), which take into account (1)
relativistic effects   
due to high electron velocities and (2) quantum effects, namely, Compton
recoil and Klein-Nishina corrections. The formula for the angle-dependent
kernel, $K(\nu,\bg{\Omega}\rightarrow\nu^\prime,\bg{\Omega^\prime})$,
is a very good approximation in the parameter range $k\te\lesssim
25$~keV and $h\nu\lesssim 50$~keV. Moreover, in the case of nonrelativistic
electrons, this formula without the temperature correction terms (eq.
[\ref{k_nr}]) gives the exact result for arbitrary photon energies,
including $h\nu\gg\me c^2$. We derived our expression in the same manner as
BR70 derived theirs, but our expression is more accurate, since it includes
the relativistic corrections consistently. 

Our expression for
$K(\nu,\bg{\Omega}\rightarrow\nu^\prime,\bg{\Omega^\prime})$ can be compared
with the corresponding formula derived by \cite{keretal86} (their eq. [41]).
Both formulae are fully algebraic and valid in roughly the same electron 
temperature range ($k\te/\me c^2\lesssim 0.1$). The expression of
\cite{keretal86} is more general than our equation (\ref{k_set}).
It holds for any $h\nu$ in all its temperature terms, while in the
case of our formula, this statement is true for the main,
nonrelativistic temperature term only. Therefore, our expression should be  
derivable from the formula of \cite{keretal86}. The referee of the present
paper informed us that both formulae indeed yield similar numerical results
for the same parameter values. On the other hand, our expression is
significantly simpler in structure and, more importantly, when
$K(\nu,\bg{\Omega}\rightarrow\nu^\prime,\bg{\Omega^\prime})$ is written in
this form, it can be integrated analytically (under some additional 
constraints on the parameters) over the scattering angle or the 
emergent photon energy. The first of these integrations leads to the
algebraic formula for the $P(\nu\rightarrow\nu^\prime)$ kernel, while the
second leads to an expression for the angular scattering function. We stress
that it is this unique integrability that makes equation (\ref{k_set}) of
interest and useful for further applications.

We derived the expression for $P(\nu\rightarrow\nu^\prime)$ by integrating
$K(\nu,\bg{\Omega}\rightarrow\nu^\prime,\bg{\Omega^\prime})$ 
over the scattering angle under the assumption $h\nu(h\nu/\me c^2)\ll 
k\te$. In this limit, it turns out to be possible to carry the term
describing the 
effect of Compton recoil out of an exponential factor that enters the
expression for $K(\nu,\bg{\Omega}\rightarrow\nu^\prime,\bg{\Omega^\prime})$
and implement the subsequent integration analytically. As a result, the 
range of applicability of our approximation for the isotropic kernel is
narrower than that of our formula for the angle-dependent kernel: $h\nu\lesssim
50$~keV, $h\nu(h\nu/\me c^2)\lesssim k\te\lesssim 25$~keV; in this range
of parameter values, the accuracy of the approximation is better than 98 per
cent. The latter constraint means that the average recoil-induced frequency
shift must be less than the typical Doppler broadening. The
opposite case, $h\nu(h\nu/\me c^2)\gtrsim k\te$, corresponds to a 
situation in which the recoil effect is predominant. In this case, the
single-scattering line profile is double-peaked, which is related to the
Rayleigh scattering phase function \citep[e.g.,][]{pozetal79,pozetal83}. It
is also worth noting that when the temperature of the matter is sufficiently
low, scattering of X-rays on neutral hydrogen and helium may become more
important than Compton scattering on free electrons \citep[see a discussion
of the recoil profile arising in this problem in][]{sunchu96}.

The paper is organized as follows. In \S 2 we report our analytical results.
For the observational astrophysicist, reading this part of the 
paper should be sufficient for finding all the information necessary for
application of the results. The main results are the formulae for the
$K(\nu,\bg{\Omega}\rightarrow\nu^\prime,\bg{\Omega^\prime})$ and
$P(\nu\rightarrow\nu^\prime)$ kernels --- equations
(\ref{k_set}),(\ref{k_nr}), and equations
(\ref{p_set}),(\ref{p_rn}),(\ref{p_k}), and (\ref{p_k_rn}), respectively. We
thoroughly examine the properties of the kernels and determine the parameter
ranges of applicability of the different approximations. In \S 2 we also:
(1) discuss the properties of the angular function for Compton scattering in
a hot plasma that results from our $K(\nu,\bg{\Omega}\rightarrow
\nu^\prime,\bg{\Omega^\prime})$ kernel; (2) show that the Fokker-Planck
expansion (to fourth order) of the kinetic equation (\ref{kinetic}) with
the kernel equation (\ref{p_set}) leads to the generalized Kompaneets equation
\citep{itoetal98,chalas98}, equation (\ref{komp_gen}); (3) derive
mildly-relativistic formulae, equations (\ref{k_ind}) and (\ref{p_ind}), for
the kernels, $K^{\rm ind}(\nu,\bg{\Omega};\nu^\prime,\bg{\Omega^\prime})$
and $P^{\rm ind}(\nu;\nu^\prime)$, that correspond to problems in
which a decisive role is played by induced Compton scattering; and (4)
give the result of the convolution of a spectrum described by the step
function with $P(\nu\rightarrow\nu^\prime)$ as an example of
application of the kernel to Comptonization problems.

The subsequent sections of the paper provide the calculation details. The
derivation procedure for the
$K(\nu,\bg{\Omega}\rightarrow\nu^\prime,\bg{\Omega^\prime})$ kernel is 
described in \S 3. The angular scattering function is calculated directly
[to a better accuracy than $K(\nu,\bg{\Omega}\rightarrow
\nu^\prime,\bg{\Omega^\prime})$] in \S 4. The formula for
the $P(\nu\rightarrow \nu^\prime)$ kernel is derived by integrating
$K(\nu,\bg{\Omega}\rightarrow\nu^\prime,\bg{\Omega^\prime})$ over the
scattering angle in \S 5. An alternative (fully independent) method for
deriving the $P(\nu\rightarrow \nu^\prime)$ kernel in the low-frequency case
($h\nu\ll k\te$) is presented in \S 6.

\section{RESULTS}

\subsection{The 
$K(\nu,\bg{\Omega}\rightarrow\nu^\prime,\bg{\Omega^\prime})$ Kernel} 

BR70 have made an attempt to take into account relativistic
effects associated with high electron velocities for the kernel
$K(\nu,\bg{\Omega}\rightarrow\nu^\prime,\bg{\Omega^\prime})$. We have
examined the derivation of these authors and found that not 
all relevant correction terms were included by them. In
particular, the relativistic corrections to the Maxwellian velocity
distribution function were neglected. The BR70 kernel is therefore of
better accuracy [correct up to terms of order $(k\te/\me c^2)^{1/2}h\nu/\me
c^2$] with respect to photon energy than with respect to electron temperature.
Therefore, this kernel is strictly only valid for nonrelativistic
electrons. In \S 3, we revise the derivation of BR70 and obtain the
following approximate formula for
$K(\nu,\bg{\Omega}\rightarrow\nu^\prime,\bg{\Omega^\prime})$, which is
correct up to terms of order $(k\te/\me c^2)^{3/2}$, $(k\te/\me
c^2)^{1/2}h\nu/\me c^2$, and $(h\nu/\me c^2)^2$:
\begin{mathletters}
\begin{eqnarray}
K(\nu,\bg{\Omega}\rightarrow\nu^\prime,\bg{\Omega^\prime})
=\nu^{-1}\frac{3}{32\pi}\sqrt{\frac{2}{\pi}}\eta^{-1/2}\,\frac
{\nu^\prime}{g}\left\{1+\mus^2+\left(\frac{1}{8}-\mus-\frac{63}{8}\mus^2+5\mus
^3\right)\eta-\frac{\mus(1+\mus)}{2}\epsilon^2
\right.
\nonumber\\
\left.
-\frac{3(1+\mus^2)}{32(1-\mus)^2}\frac{\epsilon^4}{\eta}+\mus(1-\mus^2)
\epsilon\frac{h\nu}{\me c^2}+\frac{1+\mus^2}{8(1-\mus)}\frac{\epsilon^3}
{\eta}\frac{h\nu}{\me c^2}+(1-\mus)^2\frac{h^2\nu\nu^\prime}{\me^2c^4}\right\}
\exp{\left[-\frac{\epsilon^2}{4(1-\mus)\eta}\right]},
\label{k}
\end{eqnarray}
where 
\begin{equation}
g=|\nu\bg{\Omega}-\nu^\prime\bg{\Omega^\prime}|
=(\nu^2-2\nu\nu^\prime\mus+\nu^{\prime 2})^{1/2},
\label{g}
\end{equation}
\begin{equation}
\epsilon=\frac{[2(1-\mus)]^{1/2}}{g}\left[\nu^\prime-\nu+
\frac{h\nu\nu^\prime}{\me c^2}(1-\mus)\right],
\label{epsilon}
\end{equation}
\begin{equation}
\eta=\frac{k\te}{\me c^2},
\label{eta}
\end{equation}
\label{k_set}
\end{mathletters}
and $\mus=\bg{\Omega}\bg{\Omega^\prime}$. Note that equation
(\ref{k_set}) gives the probability of a scattering event per unit
dimensionless time, $\tau=\sigma_{\rm T} N_{\rm e} ct$. 

Let us look at the sequence of terms within the braces in 
equation (\ref{k}). The main term, $1+\mus^2$, is just the Rayleigh
angular scattering function. The last term, which is proportional to
$(h\nu/\me c^2)^2$, describes the second-order Klein-Nishina
correction to the scattering cross section. The remaining five terms become
important when the electron velocities are high. These
temperature-correction terms are either incorrect or absent in the
kernel of BR70. The term of order $(h\nu/\me c^2)^2$ was not given by
BR70 either. 

Equation (\ref{k_set}) is a good approximation  (which will be supported below
by a direct comparison with results of numerical calculations) to the
kernel if both the photon energy and electron temperature are
moderately relativistic. Furthermore, this formula without the
temperature-correction terms (see the resulting eq. [\ref{k_nr}] 
below) describes the scattering of photons of arbitrary energy (including
$h\nu\gg\me c^2$) on nonrelativistic electrons ($k\te\ll \me c^2$) exactly.

The term $K(\nu,\bg{\Omega}\rightarrow\nu^\prime,\bg{\Omega^\prime})$ must obey
the detailed balance principle (including induced effects),
i.e., ensure conservation of a blackbody spectrum,
$B_{\nu}=2h\nu^3/c^2[\exp{(h\nu/k\te)}-1]^{-1}$, in thermodynamic equilibrium:
\begin{equation}
K(\nu,\bg{\Omega}\rightarrow\nu^\prime,\bg{\Omega^\prime})
\left[1+\frac{c^2B_{\nu}(\nu^\prime)}{2h\nu^{\prime3}}
\right]\frac{B_{\nu}(\nu)}{h\nu}=K(\nu^\prime,\bg{\Omega^\prime}\rightarrow\nu,
\bg{\Omega})\left[1+\frac{c^2B_{\nu}(\nu)}{2h\nu^3}
\right]\frac{B_{\nu}(\nu^\prime)}{h\nu^\prime}.
\label{k_balance1}
\end{equation}
Equation (\ref{k_balance1}) reduces to
\begin{equation}
K(\nu,\bg{\Omega}\rightarrow\nu^\prime,\bg{\Omega^\prime})=
\left(\frac{\nu^\prime}{\nu}\right)^2\exp{\left[\frac{h(\nu-\nu^\prime)}{kT_e}
\right]}K(\nu^\prime,\bg{\Omega^\prime}\rightarrow\nu,\bg{\Omega}).
\label{k_balance2}
\end{equation}
It is easily verified that our expression (\ref{k_set}) does satisfy
eq. (\ref{k_balance2}).

\subsubsection{$K(\nu,\bg{\Omega}\rightarrow\nu^\prime,\bg{\Omega^\prime})$ for
Nonrelativistic Electrons and Photons of Arbitrary Energy}   

If the electrons are nonrelativistic ($\eta\ll 1$), equation (\ref{k_set})
simplifies to
\begin{eqnarray}
K_{\rm nr}(\nu,\bg{\Omega}\rightarrow\nu^\prime,\bg{\Omega^\prime})
=\frac{3}{32\pi}\sqrt{\frac{2}{\pi}}\eta^{-1/2}\,\frac
{\nu^\prime}{\nu g}\left[1+\mus^2+(1-\mus)^2\frac{h^2\nu\nu^\prime}
{\me^2c^4}\right] 
\nonumber\\
\exp{\left\{-\frac{1}{2\eta g^2}\left[\nu^\prime-\nu
+\frac{h\nu\nu^\prime}{\me c^2}(1-\mus)\right]^2\right\}},
\label{k_nr}
\end{eqnarray}
where $g$ and $\eta$ are given by equations (\ref{g}) and (\ref{eta}),
respectively.

Since equation (\ref{k_nr}) fully takes into account the Klein-Nishina
scattering cross section, it holds true for photons of arbitrary energy when
the electrons are nonrelativistic.

If we are interested in the case in which the photons are of sufficiently
low energy, $h\nu\lesssim 0.1 \me c^2$, the second-order
Klein-Nishina correction term in equation (\ref{k_nr}) becomes small
and can be omitted. The result is
\begin{equation}
K_{\rm nr}(\nu,\bg{\Omega}\rightarrow\nu^\prime,\bg{\Omega^\prime})
=\frac{3}{32\pi}\sqrt{\frac{2}{\pi}}\eta^{-1/2}\,\frac
{\nu^\prime}{\nu g}(1+\mus^2)
\exp{\left\{-\frac{1}{2\eta g^2}\left[\nu^\prime-\nu
+\frac{h\nu\nu^\prime}{\me c^2}(1-\mus )\right]^2\right\}}.
\label{k_bab}
\end{equation}

Equation (\ref{k_bab}) includes the first-order Klein-Nishina
correction and takes into account Compton recoil; both effects are
essentially described in the exponential factor (see the derivation in \S
3). It also accounts for (to first order) the asymmetry of the scattered
profile due to the Doppler effect (the preexponential factor
$\nu^\prime/\nu$ in eq. [\ref{k_bab}] is important here), and
therefore allows one to derive the Kompaneets differential equation.

It is worth noting that the simplified derivation of the
$K(\nu,\bg{\Omega}\rightarrow\nu^\prime,\bg{\Omega^\prime})$ kernel
described in an appendix to the BR70 paper makes it possible to obtain the
exponential factor in equation (\ref{k_nr}), but not the preexponential factor
$\nu^\prime/\nu$, and therefore does not lead to the Kompaneets
equation and does not describe the energy transfer from the electrons
to the photons by scattering.

\subsubsection{Angular Scattering  Function}

By integrating the $K(\nu,\bg{\Omega}\rightarrow\nu^\prime,\bg{\Omega^\prime})$
kernel over the photon frequency upon scattering, one can determine
how many photons are scattered in a unit time through a given angle,
i.e., the angular function for scattering (defined by
eq. [\ref{ang_gen}] in \S 4):
\begin{equation}
\frac{d\sigma}{d\mus}=2\pi\int
K(\nu,\bg{\Omega}\rightarrow\nu^\prime,\bg{\Omega^\prime})\,d\nu^\prime.
\label{ang_int}
\end{equation}

We have verified, through numerical computation, that the angular
function that corresponds to the $K_{\rm nr}$ kernel given by
equation (\ref{k_nr}) is exactly the Klein-Nishina formula, describing the
cross section for Compton scattering on an electron at rest: 
\begin{equation}
\left(\frac{d\sigma}{d\mus}\right)_{\rm nr}=\frac{3}{8}\left[1+
(1-\mus)\frac{h\nu}{\me c^2}\right]^{-2}\left\{1+\mus^2+(1-\mus)^2 
\left[1+\frac{h\nu}{\me c^2}(1-\mus)\right]^{-1}\left(\frac{h\nu}{\me c^2}
\right)^2\right\}.
\label{ang_kn}
\end{equation}

\begin{figure*}[tb]
\epsfxsize=16.5cm
\epsffile[24 250 570 590]{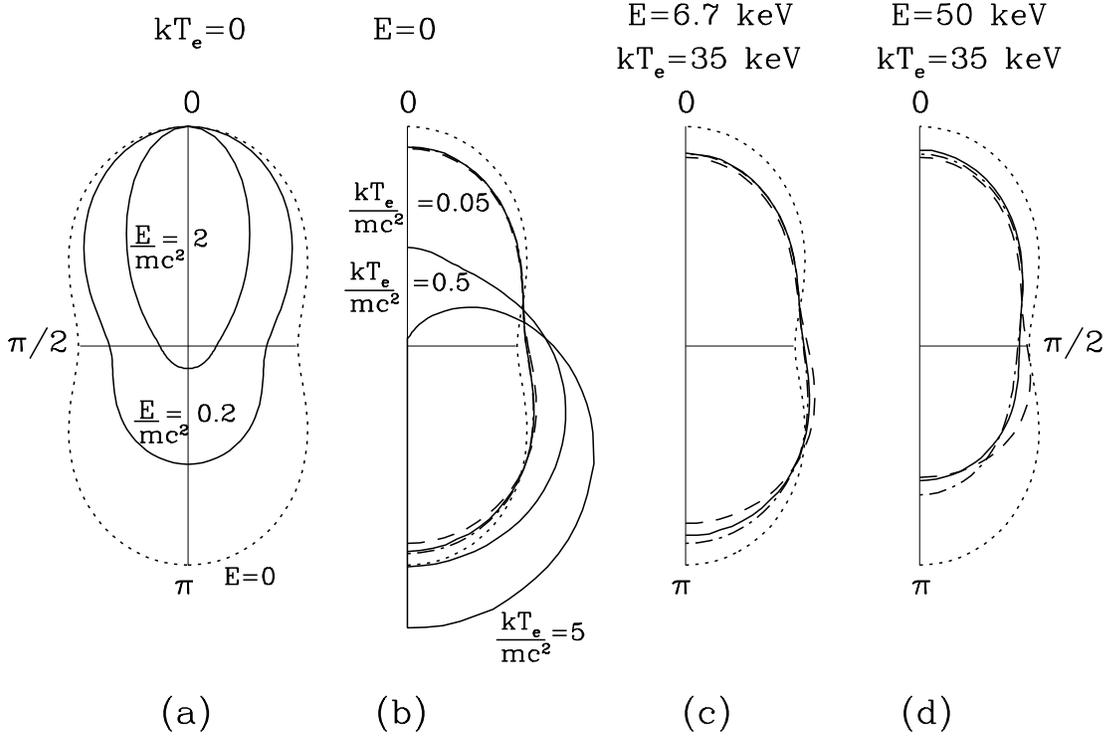}
\caption{
Angular function (in polar coordinates) for Compton
scattering on thermal electrons, in various regimes set by the values
of parameters: photon energy, $E=h\nu$, and electron temperature,
$\te$. (a) Low-temperature case: $k\te\ll h\nu$, $k\te\ll 
\me c^2$. {\sl Solid lines:} Angular functions corresponding to the
$K_{\rm nr}$ angle-dependent kernel given by eq. (\ref{k_nr}). These
patterns are described by the Klein-Nishina eq. (\ref{ang_kn}). {\sl
Dotted line:} Rayleigh angular function 
(also in the other panels). (b) Low frequency case: $h\nu\ll k\te$,
$h\nu\ll \me c^2$. {\sl Solid lines:} Results of Monte Carlo
simulations. For the mildly relativistic example, $k\te=25$~keV, also
shown are the angular function calculated from eq. (\ref{ang1}), which
results from the $K$ kernel (eq. [\ref{k_set}]) ({\sl dashed line}),
and a more accurate approximation described by eq. (\ref{ang2}) ({\sl
dash-dotted line}). (c) and (d): Both $h\nu$ and $k\te$ are mildly 
relativistic. The Monte Carlo results ({\sl solid lines}) are compared
with the results of eq. (\ref{ang1}) ({\sl dashed lines}) and eq.
(\ref{ang2}) ({\sl dash-dotted lines}).
}
\end{figure*}

Two examples of the angular function that corresponds to the
$K_{\rm nr}$ kernel and is described by equation (\ref{ang_kn}) are shown in
Figure~1a: one for the case $h\nu\ll\me c^2$ and the other for $h\nu\gg\me
c^2$. We see the well-known Klein-Nishina pattern, namely, that more
photons are scattered forward ($\mus=1$) than backward
($\mus=-1$). This angular function corresponds to the case of
nonrelativistic electrons ($k\te\ll \me c^2$).

If the electrons are mildly relativistic, the $K_{\rm nr}$ kernel becomes
inaccurate, and the temperature corrections included in the $K$
kernel (eq. [\ref{k_set}]) become important. Analytical integration of this
kernel over the scattering angle is possible if an assumption is made that
$h\nu(h\nu/\me c^2)\ll k\te\ll \me c^2$. In this limit, $K$ can be written
in the form given by equation (\ref{k_exp_set}) of \S 5, which leads to
\begin{eqnarray}
\frac{d\sigma}{d\mus}=\frac{3}{8}\left\{1+\mus^2+\left[-2(1-\mus)
(1+\mus^2)\frac{h\nu}{\me c^2}+(1-\mus)^2(4+3\mus^2)
\left(\frac{h\nu}{\me c^2}\right)^2...\right]
\right.
\nonumber\\
\left.
+2(1-2\mus-3\mus^2+2\mus^3)\frac{k\te}{\me c^2}
+O((h\nu/\me c^2)(k\te/\me c^2),(k\te/\me c^2)^2)\right\}.
\label{ang1}
\end{eqnarray}

The expression in square brackets in equation (\ref{ang1}) is an expansion   
series in powers of $h\nu/\me c^2$ resulting from equation (\ref{ang_kn})
(only two leading terms are presented). More interesting is the
correction term of order $k\te/\me c^2$. The origin of this term is
related to Doppler aberration and has nothing to do with quantum
effects. The terms that are given in implicit 
form in equation (\ref{ang1}), i.e. $O(...)$, indicate the order of
inaccuracy of our approximation for the kernel. Although equation
(\ref{ang1}) was obtained under the assumption that
$h\nu(h\nu/\me c^2)\ll k\te\ll \me c^2$, it holds true
for arbitrary proportions of $h\nu$ and $k\te$, which we have
verified by a direct numerical calculation of the integral (\ref{ang_int}).

It is also possible to directly calculate the angular function, not using the
$K(\nu,\bg{\Omega}\rightarrow \nu^\prime,\bg{\Omega^\prime})$ 
kernel. The corresponding derivation procedure, which is similar to 
but simpler than that for $K(\nu,\bg{\Omega}\rightarrow
\nu^\prime,\bg{\Omega^\prime})$, is described in \S 4. The
result, which is more accurate than equation (\ref{ang1}), is
\begin{eqnarray}
\frac{d\sigma}{d\mus}=\frac{3}{8}\left[1+\mus^2-2(1-\mus)
(1+\mus^2) 
\frac{h\nu}{\me c^2}+2(1-2\mus-3\mus^2+2\mus^3)
\frac{k\te}{\me c^2}
\right.
\nonumber\\
\left.
+(1-\mus)^2(4+3\mus^2)\left(\frac{h\nu}{\me c^2}\right)^2+
(1-\mus)(-7+14\mus+9\mus^2-10\mus^3)\frac{h\nu}{\me c^2}\frac{k\te}
{\me c^2} 
\right.
\nonumber\\
\left.
+(-7+22\mus+9\mus^2-38\mus^3+20\mus^4)\left(\frac{k\te}{\me c^2}\right)^2 
+...\right].
\label{ang2}
\end{eqnarray}
We see that the first-order correction terms, $O(h\nu/\me c^2)$ and
$O(k\te/\me c^2)$, and the second-order term proportional to
$(h\nu/\me c^2)^2$ are the same as in equation (\ref{ang1}). The last two
terms in equation (\ref{ang2}), $O((h\nu/\me c^2)(k\te/\me c^2))$ and
$O((k\te/\me c^2)^2)$, belong to the next-order approximation
(with respect to $K$ given by eq. [\ref{k_set}]) to the kernel.

Integration of equation (\ref{ang2}) over all scattering angles leads to the
well-known expression that describes the total cross section for
Compton scattering on Maxwellian electrons [defined as
$\sigma=(\lambda N_{\rm e})^{-1}$, where $\lambda$ is the photon mean
free path] in the mildly relativistic limit
\citep[e.g.,][]{pozetal83,sheetal88}:
\begin{equation}
\sigma=\sigma_{\rm T}\left[1-2\frac{h\nu}{\me c^2}-5\frac{h\nu}{\me c^2}
\frac{k\te}{\me c^2}+\frac{26}{5}\left(\frac{h\nu}{\me c^2}
\right)^2\right].
\label{sigma}
\end{equation}
Note that the pure temperature terms, $O((k\te/\me c^2)^n)$, in
equation (\ref{ang2}) give no contribution to $\sigma$. This is a
well-known fact, which means that the total cross section for
low-energy photons in a hot plasma is equal to the Thomson
cross section. At the same time, the angular function
(eq. [\ref{ang2}]) is different from the Rayleigh function.

In Figures~1b--1d, several examples of the angular function for Compton
scattering on hot electrons are presented; the results of Monte Carlo
simulations are compared with the results of the calculation by the
approximate formulae (\ref{ang1}) and (\ref{ang2}).

Figure~1b demonstrates scattering of low-frequency photons: $h\nu\ll
k\te$, $h\nu\ll \me c^2$. The angular function in this case
is totally different from the Klein-Nishina one --- compare with Figure~1a.
Let us first consider the mildly relativistic temperature range $k\te\lesssim
25$~keV, within which equation (\ref{ang1}) (with the terms containing
$h\nu/\me c^2$ vanishing) is a good approximation. Scattering is
somewhat suppressed (compared to the Rayleigh angular function, which
corresponds to the $K_{\rm nr}$ kernel in this case) both in the forward
and backward directions. There is, however, a noticeable enhancement
in the number of photons scattered through intermediate angles, between
$69^\circ$ and $138^\circ$; these values are found by equating the
correction term of order $k\te/\me c^2$ in equation (\ref{ang1}) to
zero. The temperature-relativistic correction to the Rayleigh angular
function reaches a maximum of $12(\eta/0.05)$\% at an angle of
$105^\circ$. The relativistic reduction of the angular function is
maximal, $10(\eta/0.05)$\%, for the two extreme values of the
scattering angle, $0$ and $\pi$. It is evident from Figure~1b that the
approximation represented by equation (\ref{ang2}) is more accurate that that
given by equation (\ref{ang1}). The inclusion of the correction term of order
$(k\te/\me c^2)^2$ is particularly important for very large
scattering angles (close to $\pi$). Equation (\ref{ang2}) proves to
be a good approximation to the actual scattering angular function  in
the range $k\te\lesssim 35$~keV.

As the temperature becomes significantly relativistic (see the patterns
for $k\te=0.5 \me c^2$ and $5 \me c^2$ in Fig.~1b),
the scattering angular function modifies further and becomes totally 
unlike the Rayleigh angular function. Only the results of Monte Carlo
simulations are shown for these cases, because the convergence of the
expansion series in powers of $k\te/\me c^2$ for the
angular function becomes poor when $k\te$ exceeds $\sim
40$~keV. Scattering in the forward direction is now heavily suppressed
(i.e., the plasma effectively screens itself from the radiation
incident from outside), while more and more photons are scattered
through angles between $\pi/2$ and $\pi$. In particular, at temperatures
$k\te\gtrsim 0.5 \me c^2$, the number of photons scattered
through an angle of $\pi$ is higher than in the case $k\te=0$. In the
limit $k\te\gg \me c^2$, the angular function is described by the law
$d\sigma/d\mus=(1-\mus)/2$ \citep{sazsun00}.

Figures~1c and 1d show two examples of the angular scattering function
when both $h\nu$ and $k\te$ are mildly relativistic. The
approximations described by equations (\ref{ang1}) and (\ref{ang2})
work well for photon energies $h\nu\lesssim 50$~keV within the
temperature ranges quoted above. 

The phenomenon of anisotropic (backward) Compton scattering by a hot
plasma has been previously mentioned in astrophysical literature
\citep{ghietal91,titarchuk94,pousve96,gieetal99}. In particular,
\cite{haardt93} has performed a semianalytical calculation (using
Monte Carlo simulations) of the angular scattering function, as well
as other related quantities, and obtained results for the case of
low-frequency radiation ($h\nu\ll \me c^2$) that are similar to those
depicted in Figure~1b.

\subsubsection{The Moments of the Kernel}

Using equation (\ref{k_exp_set}) of \S 5 it is possible to
calculate the moments of the
$K(\nu,\bg{\Omega}\rightarrow\nu^\prime,\bg{\Omega^\prime})$ kernel,
in a similar way as we derived above the angular function. Here we define the
moments as
\begin{equation}
\langle(\Delta\nu)^n\rangle_{\mus}=2\pi\int
K(\nu,\bg{\Omega}\rightarrow\nu^\prime,\bg{\Omega^\prime})
(\nu^\prime-\nu)^n d\nu^\prime,
\label{mom2}
\end{equation}
so that the integration of $\langle(\Delta\nu)^n\rangle_{\mus}$ over
$\mus$ gives the moments of the $P(\nu\rightarrow\nu^\prime)$ kernel
for the isotropic problem, which were defined in equation (\ref{mom1}).

The result for the first four moments is
\begin{eqnarray}
\langle\Delta\nu\rangle_{\mus} &=&
\frac{3\nu}{8}\left[(1-\mus)(1+\mus^2)\left(4\frac{k\te}{\me c^2}
-\frac{h\nu}{\me c^2}\right)+2(7-21\mus+5\mus^2+19\mus^3-10\mus^4)
\left(\frac{k\te}{\me c^2}\right)^2
\right.
\nonumber\\
& &\left.+\,\frac{-37+81\mus-65\mus^2+41\mus^3-20\mus^4}{2}\frac{h\nu}
{\me c^2}\frac{k\te}{\me c^2}+3(1-\mus)^2(1+\mus^2)
\left(\frac{h\nu}{\me c^2}\right)^2
\right],
\nonumber\\
\langle(\Delta\nu)^2\rangle_{\mus} &=&
\frac{3\nu^2}{8}\left[2(1-\mus)(1+\mus^2)\frac{k\te}{\me c^2}
+(37-81\mus+65\mus^2-41\mus^3+20\mus^4)\left(\frac{k\te}{\me c^2}\right)^2
\right.
\nonumber\\
& &\left.-18(1-\mus)^2(1+\mus^2)\frac{h\nu}{\me c^2}\frac{k\te}{\me c^2}
+(1-\mus)^2(1+\mus^2)\left(\frac{h\nu}{\me c^2}\right)^2
\right], 
\nonumber\\
\langle(\Delta\nu)^3\rangle_{\mus} &=& \frac{3\nu^3}{8}(1-\mus)^2(1+\mus^2) 
\left[36\left(\frac{k\te}{\me c^2}\right)^2-6\frac{h\nu}{\me c^2}\frac{k\te} 
{\me c^2}\right], 
\nonumber\\
\langle(\Delta\nu)^4\rangle_{\mus} &=& \frac{3\nu^4}{8}12(1-\mus)^2(1+\mus^2)
\left(\frac{k\te}{\me c^2}
\right)^2. 
\label{k_moments}
\end{eqnarray}
The moments of higher degrees turn out to be at least of order $\eta^3$. The
importance of including relativistic corrections in the kernel
becomes clear from examining the terms proportional to $\eta^2$ and $\eta
h\nu/\me c^2$ in the expressions for the first two moments, which become
comparable to the main terms already at moderate $\eta, h\nu/\me c^2\sim 0.02$
for large scattering angles. The first and second moments describe
correspondingly the average frequency increment by scattering and the
broadening of the scattered profile.

It should be noted that the accuracy, to within
$O(\eta^2,\eta h\nu/\me c^2,(h\nu/\me c^2)^2)$, of
the derived moments is better than the accuracy, to within
$O(\eta^{3/2}, \eta^{1/2}h\nu/\me c^2)$, of the
kernel itself. The reason for this is that
$K(\nu,\bg{\Omega}\rightarrow\nu^\prime,\bg{\Omega^\prime})$ is
multiplied by a small quantity, the frequency shift
($\sim\nu\eta^{1/2}$) raised to a certain power, when the moments are
calculated.

\subsubsection{Comparison of the Analytical Formulae for the Kernel
with Numerical Results}

We performed a series of Monte Carlo simulations (a modification
of the code described by Pozdnyakov et al. 1983 was used) to
evaluate the accuracy of our analytical expressions for the
$K(\nu,\bg{\Omega}\rightarrow\nu^\prime,\bg{\Omega^\prime})$
kernel. Figures~2--7 show examples of spectra that may 
form through single scattering (with a given angle) of a monochromatic line
on thermal electrons in various scattering regimes that are set by the
values of the parameters: $k\te$, $h\nu$, $\mus$. The
numerical results are compared with the analytical kernels $K$
(mildly relativistic) and $K_{\rm nr}$ (non-relativistic).

As inferred from Figures~2--7 and further from the entire set of
results of our numerical calculations, the mildly relativistic equation
(\ref{k_set}) is a very good approximation for electron temperatures
$k\te\lesssim 25$~keV and photon energies $h\nu\lesssim 50$~keV. In
this range of parameter values, the accuracy is better than 98\%,
except in the far wings of the scattered profile. The latter can
be roughly defined as the regions where $|\epsilon|\gtrsim
0.5(1-\mus)^{1/2}$ (this size should be compared with the
characteristic width of the line, which is much smaller:
$[(1-\mus)\eta]^{1/2}$). For very large scattering angles,
$\mus\lesssim -0.8$, relativistic corrections are particularly 
important. In this angular range, the accuracy quoted above is
achieved in a more narrow temperature range, $k\te\lesssim 15$~keV.

\begin{figure*}[b]
\epsfxsize=16.5cm
\epsffile[-110 190 685 700]{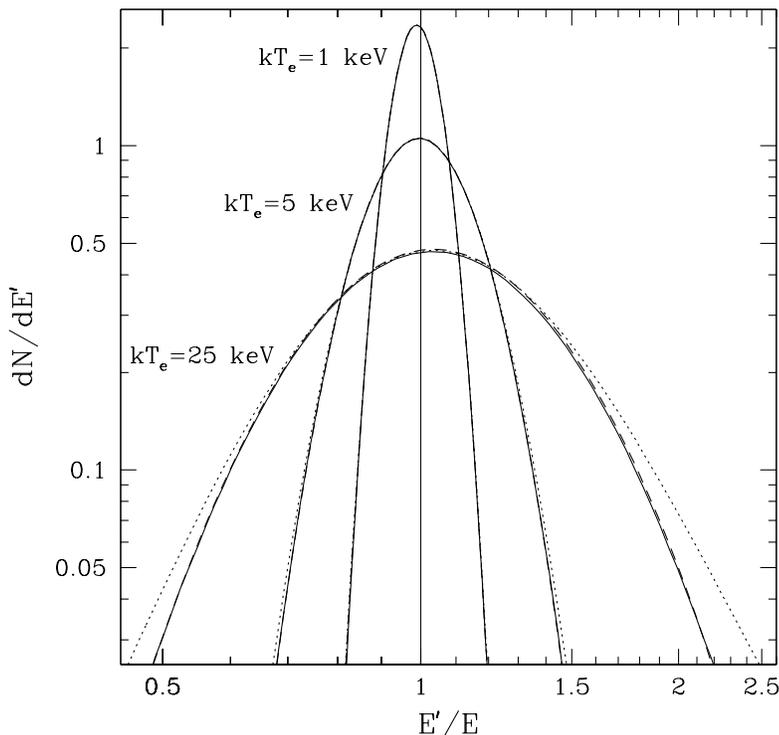}
\caption{
Photon number spectra resulting from the single scattering of a
monochromatic line of energy, $h\nu=6.7$~keV, on hot electrons with a
scattering angle of $\pi/2$, for a number of values of electron temperature.
The results of Monte Carlo simulations ({\sl solid lines}) are compared
with various approximations for the angle-dependent kernel: $K_{\rm nr}$
(eq. [\ref{k_nr}], {\sl dotted lines}) and $K$ (eq. [\ref{k_set}], {\sl
dashed lines}). Note the increasing influence of relativistic
corrections (included in the $K$ kernel) on the spectrum as the
electron temperature becomes higher.
}
\end{figure*}

\begin{figure*}[tb]
\epsfxsize=16.5cm
\epsffile[-110 190 685 700]{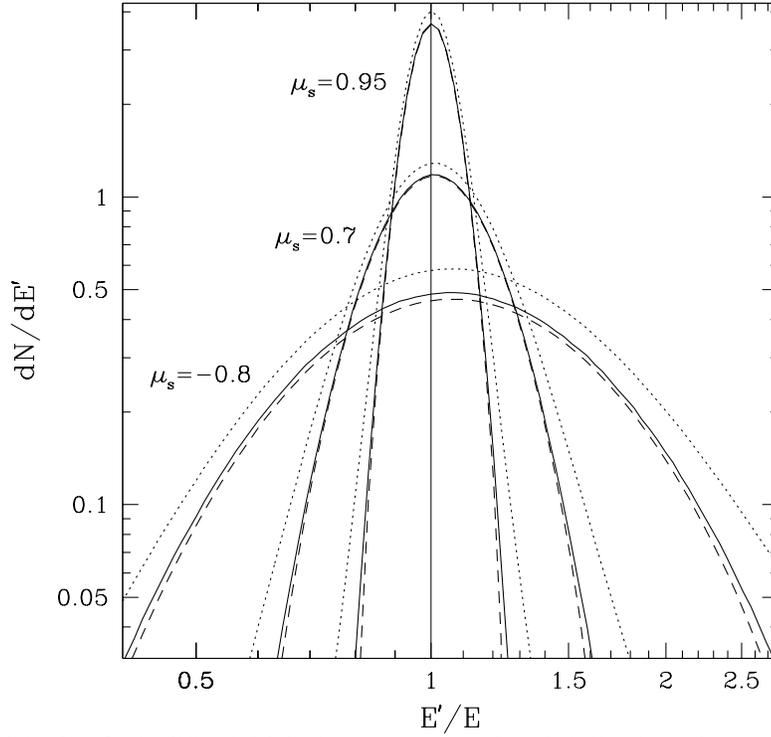}
\caption{
Same as Fig.~2, but for a fixed, mildly relativistic
electron temperature, $k\te=25$~keV, and a varying scattering
angle. Note that increasing the scattering angle for a given 
temperature has a similar broadening effect on the spectrum as increasing the
temperature with the scattering angle fixed (compare with Fig.~2). 
}
\end{figure*}

\begin{figure*}[tb]
\epsfxsize=16.5cm
\epsffile[-110 190 685 700]{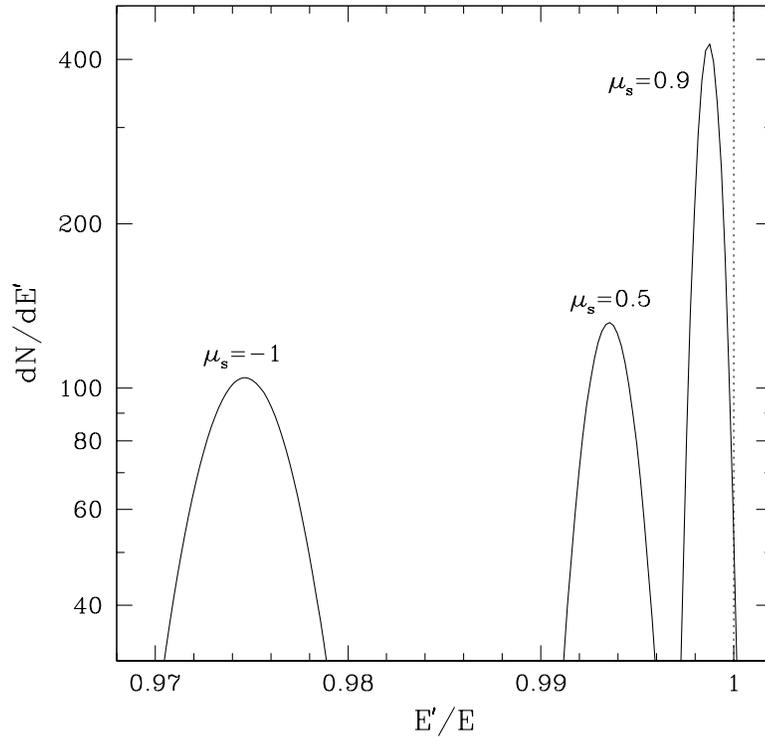}
\caption{
Spectra resulting from the single scattering of a monochromatic
line of energy, $h\nu=6.7$~keV, on low-temperature electrons,
$\te=10^4$~K, for a number of scattering angles. The corrections
relevant for relativistic electrons are not important in this case. The
presented spectra were calculated from eq. (\ref{k_nr}) for the
$K_{\rm nr}$ kernel, which gives the exact result in the case of 
nonrelativistic electrons and leads to the same results as Monte Carlo
simulations. 
}
\end{figure*}

\begin{figure*}[tb]
\epsfxsize=16.5cm
\epsffile[-110 190 685 700]{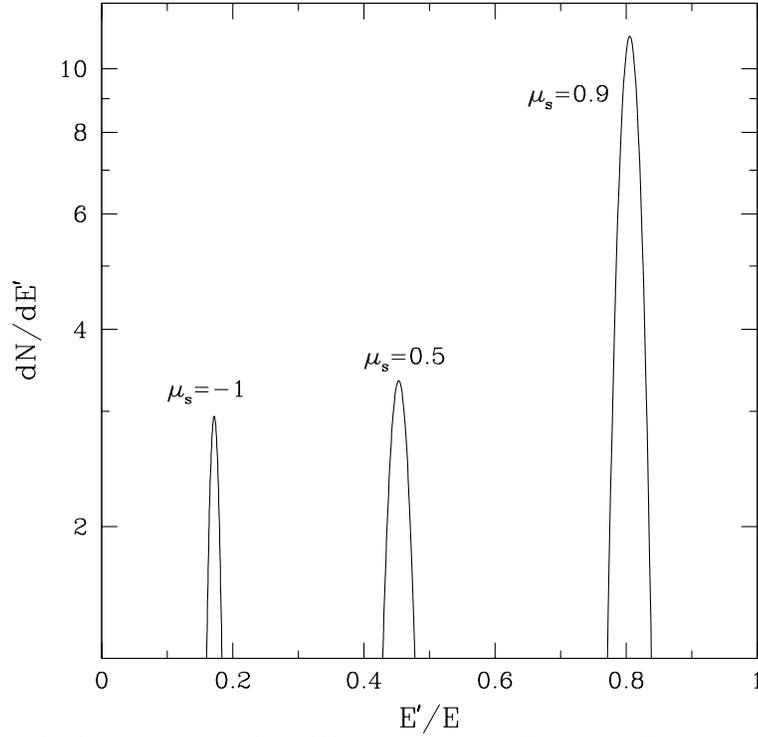}
\caption{
Same as Fig.~4, but for high-energy photons,
$h\nu=1238$~keV, and electrons of $k\te=1$~keV. The $K_{\rm nr}$ kernel
given by eq. (\ref{k_nr}) is still very accurate at this electron
temperature and leads to practically identical results with those of
Monte Carlo simulations. Note that the total number of scattered photons is
drastically decreased on going from $\mus=0.5$ to $\mus=-1$. This is due to
the increasing effect of Klein-Nishina corrections on the scattering
cross section (compare with the nonrelativistic case in Fig.~4).
}
\end{figure*}

\begin{figure*}[tb]
\epsfxsize=16.5cm
\epsffile[-110 190 685 700]{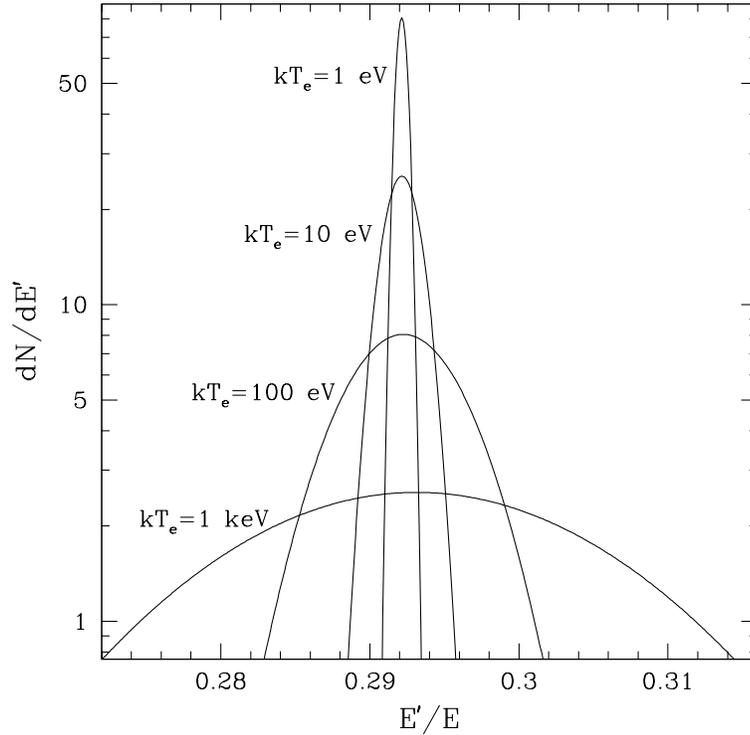}
\caption{
Same as Fig.~5 ($h\nu=1238$~keV), but for a fixed
scattering angle, $\mus=0$, and a varying, nonrelativistic electron
temperature. 
}
\end{figure*}

\begin{figure*}[tb]
\epsfxsize=16.5cm
\epsffile[-110 190 685 700]{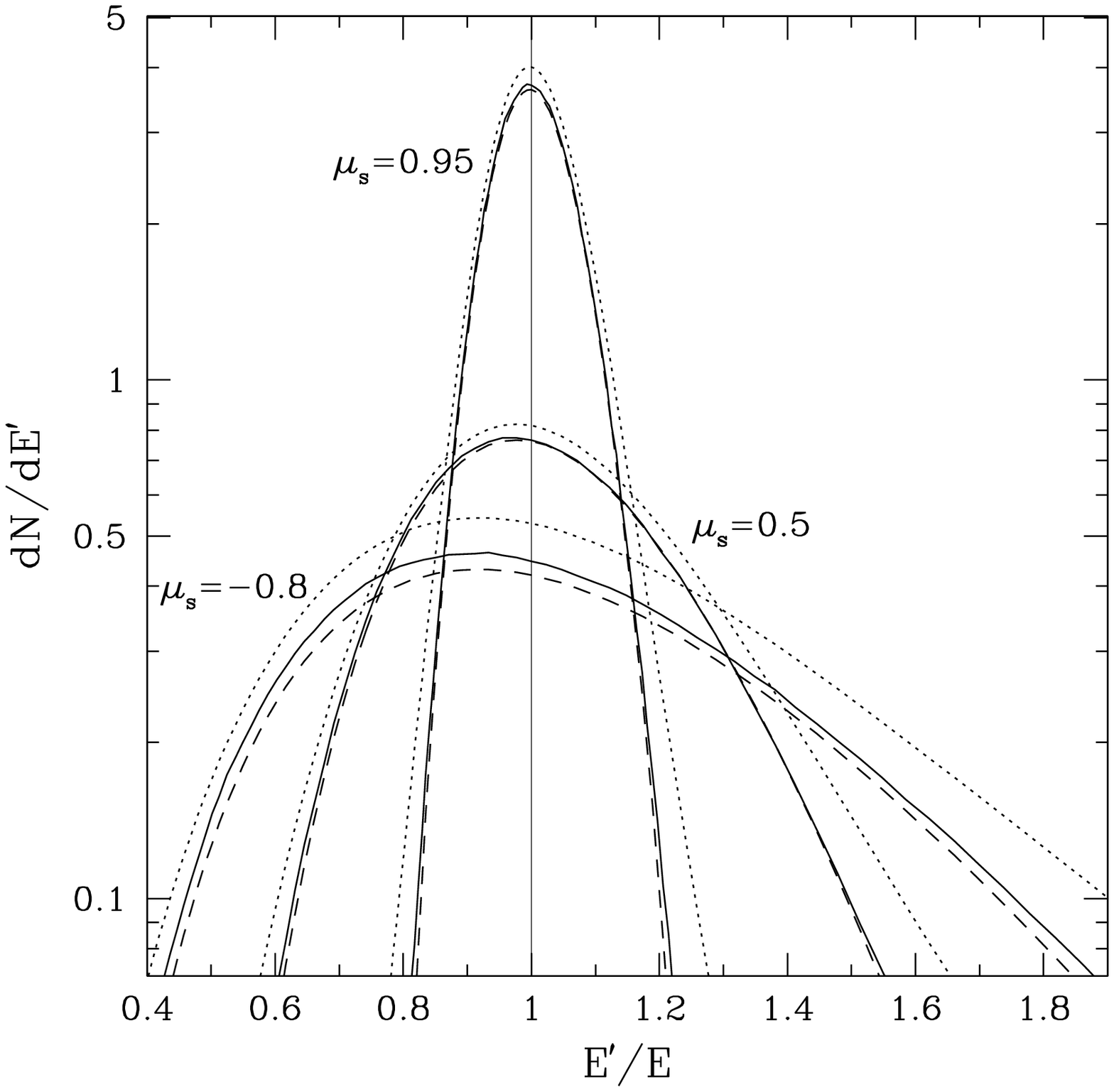}
\caption{
Spectra forming through single scattering of monochromatic
radiation, $h\nu=50$~keV, on mildly relativistic thermal electrons, 
$k\te=25$~keV, for a set of scattering angles. The results of
Monte Carlo simulations ({\sl solid lines}) are compared with the
different approximations for the angle-dependent kernel: $K_{\rm nr}$
(eq. [\ref{k_nr}], {\sl dotted lines}) and $K$ (eq. [\ref{k_set}], {\sl
dashed lines}). Temperature relativistic corrections, which are
included in the $K$ kernel but not in the $K_{\rm nr}$ kernel, are of 
importance in this case.
}
\end{figure*}

One can safely use the $K_{\rm nr}$ kernel when $k\te\lesssim 5$~keV. The
photon energy can take arbitrary values in this case because equation
(\ref{k_nr}) is exact for nonrelativistic electrons, as we 
pointed out before. This is demonstrated by Figures~5--7, which illustrate
the scattering of a line with $h\nu=1238$~keV (which
corresponds to one of the strongest $\gamma$ lines produced through 
radioactive decay of $^{56}$Co during supernova explosions) by thermal
electrons, for a number of values of the electron temperature and
scattering angle. Note that even for a temperature of $\te=10^4$~K, the
Doppler width of the scattered line is $\sim 1$~keV for a scattering angle
of $\pi/2$. The Ge cryogenic detectors of the {\sl International
Gamma-Ray Astrophysical Observatory} ({\sl INTEGRAL}), scheduled for launch
in 2002, will be capable of measuring such broadening. 

The line broadening arising from the Doppler effect depends on two
parameters: the electron temperature, $k\te$, and the scattering angle, $\mus$.
The width of the profile as calculated from equation (\ref{k_nr}) is
strictly proportional to $[(1-\mus)\eta]^{1/2}$. Therefore, in the
nonrelativistic limit, one can only determine this combination of the two
parameters, rather than $k\te$ and $\mus$ separately, from a measurement of
line broadening. This point is illustrated in Figures~2 and 3. One can see
that identical profiles can be obtained by varying either the temperature or
the scattering angle. However, if the electrons are mildly relativistic, the
correction terms in equation (\ref{k_set}) become important, with their own
dependence on $\mus$. This property makes it possible, in principle, to find
both $k\te$ and $\mus$ from a measurement of a scattered line. However, in
many situations, it would be easier to determine $\mus$ by measuring the
recoil-induced shift of the line (i.e., the position of the line on the
frequency axis, see Figs.~2--7) which is proportional to $ (1-\mus)h\nu/\me
c^2$.

\subsection{The $P(\nu\rightarrow \nu^\prime)$ Kernel for the
Isotropic Problem}

In the preceeding paragraph we presented the analytical formula
for the $K(\nu,\bg{\Omega}\rightarrow\nu^\prime,\bg{\Omega^\prime})$
kernel, which can be used for describing spectra forming as a result of single
Compton scattering of a monochromatic line, with a given angle between the
direction from which photons are supplied ($\bg{\Omega}$) and the
observer's direction ($\bg{\Omega^\prime}$). However, in most astrophysical
situations we are presented with some angular distribution of incident
radiation. In such cases, one needs to convolve the
$K(\nu,\bg{\Omega}\rightarrow\nu^\prime,\bg{\Omega^\prime})$ kernel with
this initial distribution in order to determine the emergent spectrum.

If the initial distribution is isotropic, we arrive at the kinetic
equation (\ref{kinetic}), with the kernel derivable by the integration
given in equation (\ref{k_int}). This integral can be performed
analytically for the 
$K(\nu,\bg{\Omega}\rightarrow\nu^\prime,\bg{\Omega^\prime})$ kernel 
given by equation (\ref{k_set}) if we assume that $h\nu(h\nu/\me
c^2)\ll k\te$, so that the Doppler broadening prevails over the recoil shift.
The details of the integration procedure are given in \S 5. The essence of
this calculation is to write the argument of the exponential in equation
(\ref{k_set}) as a trinomial. The two terms dependent on the photon energy
($h\nu$) then turn out to be small with respect to the main term
describing the Doppler broadening (because of our assumption that
$h\nu\sim k\te$), and therefore can be carried out of the
exponential. The subsequent integration of $K$ over the scattering
angle becomes straightforward. The result is
\begin{mathletters}
\begin{eqnarray}
P(\nu\rightarrow\nu^\prime)=\nu^{-1}\sqrt{\frac{2}{\pi}}\eta^{-1/2}
\left\{\left[1+\sqrt{2}\delta\left(1-\frac{h\nu}{k\te}\right)\eta^{1/2}
-4\delta^2\frac{h\nu}{\me c^2}
\right.\right.
\nonumber\\
\left.\left.
+2\sqrt{2}\delta^3\left(-2+\frac{1}{3}\left(\frac{h\nu}
{k\te}\right)^2\right)\frac{h\nu}{\me c^2}\eta^{1/2}\right]p_0
+\left[1+\sqrt{2}\delta\left(1-\frac{h\nu}{k\te}\right)\eta^{1/2}\right]p_t 
\right.
\nonumber\\
\left.
+\left[1+\sqrt{2}\delta\left(3-\frac{h\nu}{k\te}\right)\eta^{1/2}
\right]p_r \right\},
\label{p}
\end{eqnarray}
where
\begin{eqnarray}
p_0 &=& \left(\frac{11}{20}+\frac{4}{5}\delta^2+\frac{2}{5}\delta^4\right)F
+|\delta|\left(-\frac{3}{2}-2\delta^2-\frac{4}{5}\delta^4\right)G,
\nonumber\\
p_t &=& \left[\left(-\frac{1091}{1120}-\frac{507}{560}\delta^2
+\frac{57}{35}\delta^4+\frac{68}{35}\delta^6\right)F+|\delta|\left(\frac{9}{4}
+\delta^2-\frac{26}{5}\delta^4-\frac{136}{35}\delta^6\right)G\right]
\eta,
\nonumber\\
p_r &=& \left[\left(-\frac{23}{280}+\frac{26}{35}\delta^2+\frac{34}{35}
\delta^4+\frac{16}{35}\delta^6\right)F+|\delta|\left(-2\delta^2-\frac{12}{5}
\delta^4-\frac{32}{35}\delta^6\right)G\right]
\left(\frac{h\nu}{\me c^2}\right)^2\eta^{-1},
\nonumber\\
F &=& \exp{(-\delta^2)},\,\,\,G=\int_{|\delta|}^{\infty}\exp{(-t^2)}\,dt=0.5
\pi^{1/2}Erfc(|\delta|),\,\,\,\delta=(2\eta)^{-1/2}
\frac{\nu^\prime-\nu}
{\nu^\prime+\nu}.
\label{pp}
\end{eqnarray}
\label{p_set}
\end{mathletters}

The kernel given by equation (\ref{p_set}) is a series in powers of
$\eta^{1/2}$ (given that $h\nu\sim k\te$), written up to the
third order.

In the case of scattering of low-frequency radiation in a hot
plasma ($h\nu\ll k\te$), equation (\ref{p_set}) simplifies significantly:
\begin{eqnarray}
P_T=\nu^{-1}\sqrt{\frac{2}{\pi}}\eta^{-1/2}\left[1+\sqrt{2}\delta\eta^{1/2}
\right](p_0+p_t).
\label{pt}
\end{eqnarray}
Equation (\ref{pt}), which totally neglects Compton recoil, can be
obtained independently from the
$K(\nu,\bg{\Omega}\rightarrow\nu^\prime,\bg{\Omega^\prime})$ kernel,
using a simple calculation procedure that employs a transition to the rest
frame of the scattering electron (\S 6). As a matter of fact, we first
obtained equation (\ref{pt}), and this formula allowed us to check (by applying
the detailed balance principle; see below) some of the terms
dependent on $h\nu/\me c^2$ in the more general equation (\ref{p_set}).

We have verified that the expression (\ref{p_set}) obeys the detailed
balance principle, satisfying (in all orders up to $\eta^{3/2}$) the equation
\begin{equation}
P(\nu\rightarrow\nu^\prime)=
\left(\frac{\nu^\prime}{\nu}\right)^2\exp{\left[\frac{h(\nu-\nu^\prime)}
{k\te}\right]}P(\nu^\prime\rightarrow\nu),
\label{p_balance}
\end{equation}
which is the analog of the corresponding equation for the
$K(\nu,\bg{\Omega}\rightarrow\nu^\prime,\bg{\Omega^\prime})$ kernel
(eq. [\ref{k_balance2}]).

\subsubsection{The Kernel Leading to the Kompaneets Equation}

The expression (\ref{p_set}) is the sum of four leading terms of the
series in powers of $\eta^{1/2}$ for the $P(\nu\rightarrow\nu^\prime)$
kernel. By retaining a smaller number of terms in this series, one can
build up cruder approximations to the kernel. Below, we shall consider two such
approximations.

The least accurate approximation uses the first term in the series
given in equation (\ref{p}): 
\begin{equation}
P_0(\nu\rightarrow\nu^\prime)=\nu^{-1}\sqrt{\frac{2}{\pi}}\eta^{-1/2}p_0,
\label{p_0}
\end{equation}  
with $p_0$ given by equation (\ref{pp}).

The expression (\ref{p_0}) is equivalent to the formula obtained by
\cite{hummih67}. $P_0$ is symmetric in frequency shift:
$p_0(-\delta)=p_0(\delta)$. Therefore, it only describes the Doppler
(random-walk) broadening, not accounting for the average increase in
the photon energy by the Doppler effect.

A more accurate approximation can be obtained by summing up two leading
terms in the series given in equation (\ref{p}): 
\begin{equation}
P_K(\nu\rightarrow\nu^\prime)=\nu^{-1}\sqrt{\frac{2}{\pi}}\eta^{-1/2}\left[1+
\sqrt{2}\delta\left(1-\frac{h\nu}{k\te}\right)\eta^{1/2}\right]p_0.
\label{p_k}
\end{equation}

The $P_K$ kernel is necessary and sufficient for deriving the Kompaneets
equation (\ref{komp}). For this reason, we refer to it as the
``Kompaneets equation kernel''. When $h\nu\ll k\te$, equation (\ref{p_k}) is
equivalent to the formula obtained by \cite{sunyaev80}. The $P_K$ kernel
already accounts for the asymmetry of the scattered line and the
corresponding photon heating. The Kompaneets equation kernel also
takes into account (to first order) Compton recoil.

\subsubsection{The Moments and Normalization of the Kernel}

Important information on the properties of the
$P(\nu\rightarrow\nu^\prime)$ kernel is provided by its moments, which
were defined in equation (\ref{mom1}).

The required integration is readily performed for the kernel given by
equation (\ref{p_set}) after we express the differential $d\nu^\prime$
appearing in equation (\ref{mom1}) through $d\delta$,
\begin{equation}
d\nu^\prime=2\sqrt{2}\nu\eta^{1/2}(1+2\sqrt{2}\eta^{1/2}\delta+6\delta^2\eta
+8\sqrt{2}\delta^3\eta^{3/2}+...)\,d\delta.
\end{equation}
The quantity $\delta$ was introduced in equation (\ref{pp}). The 
subsequent integration over $\delta$ should be carried out in the limits
$-\infty$ to $\infty$.

Collecting terms up to
$O(\eta^2,\eta h\nu/\me c^2,(h\nu/\me c^2)^2)$, we
obtain 
\begin{eqnarray}
\langle\Delta\nu\rangle &=& \nu\left[4\frac{k\te}{\me c^2}
-\frac{h\nu}{\me c^2}+10\left(\frac{k\te}{\me
c^2}\right)^2-\frac{47}{2} \frac{h\nu}{\me c^2}\frac{k\te}{\me c^2}
+\frac{21}{5}\left(\frac{h\nu}{\me c^2} 
\right)^2\right],
\nonumber\\
\langle(\Delta\nu)^2\rangle &=& \nu^2\left[2\frac{k\te}{\me c^2}+47
\left(\frac{k\te}{\me c^2}\right)^2-\frac{126}{5}\frac{h\nu}{\me
c^2} \frac{k\te}{\me c^2}+\frac{7}{5}\left(\frac{h\nu}{\me
c^2}\right)^2\right],
\nonumber\\
\langle(\Delta\nu)^3\rangle &=& \nu^3\left[\frac{252}{5}\left(\frac{k\te}
{\me c^2}\right)^2-\frac{42}{5}\frac{h\nu}{\me c^2}\frac{k\te}{\me c^2}
\right],
\nonumber\\
\langle(\Delta\nu)^4\rangle &=& \nu^4\left[\frac{84}{5}\left(\frac{k\te}
{\me c^2}\right)^2\right].
\label{p_moments}
\end{eqnarray}

The moments of higher degrees turn out to be at least of order
$\eta^3$. The values for the moments above reproduce the results of
\cite{itoetal98,chalas98} (more general expressions for the first two
moments are also available; see Shestakov et al. 1988; \S 4.3 in
Nagirner \& Poutanen 1994). It is important to mention that the equations
(\ref{p_moments}) are valid for arbitrary values of the $h\nu/k\te$
ratio, including the case $k\te=0$, in contrast to the kernel equation
(\ref{p_set}) itself, which holds in the limit $h\nu(h\nu/\me
c^2)\lesssim k\te$ only (see the next paragraph for a discussion of the
scope of the analytical approximations).

The moments (\ref{p_moments}) can also be obtained by integrating over the
scattering angle the moments of the angle-dependent
$K(\nu,\bg{\Omega}\rightarrow \nu^\prime,\bg{\Omega^\prime})$ kernel (eq.
[\ref{k_moments}]).

The normalization of the kernel equation (\ref{p_set}) can be
calculated similarly to its moments:
\begin{equation}
\int P\,d\nu^\prime=1-2\frac{h\nu}{\me
c^2}+\left[-\frac{53}{10}\left(\frac {k\te}{\me
c^2}\right)^2-\frac{44}{5}\frac{h\nu}{\me c^2}\frac{k\te}{\me
c^2} +\frac{63}{20}\left(\frac{h\nu}{\me c^2}\right)^2\right].
\label{p_norm}
\end{equation}

This expression should be compared with the known expansion series for
the total cross section, equation (\ref{sigma}). One can see that the
terms in square brackets in equation (\ref{p_norm}) describe the
inaccuracy of the approximation in equation (\ref{p_set}), whereas the
term $-2h\nu/\me c^2$ 
corresponds to the actual first-order Klein-Nishina correction to the
cross section (see eq. [\ref{sigma}]). Note that equation (\ref{p_norm})
does not describe the second-order Klein-Nishina correction to the scattering
cross section, although our expression for the angle-dependent kernel
$K(\nu,\bg{\Omega}\rightarrow\nu^\prime,\bg{\Omega^\prime})$ (eq.
[\ref{k_set}]) does contain the corresponding term. This term was omitted when
we integrated $K(\nu,\bg{\Omega}\rightarrow\nu^\prime,\bg{\Omega^\prime})$
over the scattering angle (in \S 5), because in the current limit ($h\nu\sim
k\te$), inclusion of this term in the formula for
$P(\nu\rightarrow\nu^\prime)$ (eq. [\ref{p_set}]) would be inconsistent
with the absence of some (unknown) terms of the same order, such as
$O(\eta^2)$ or $O((h\nu/\me c^2)^3\eta^{-1})$, in this formula.

For the zero-order kernel, $P_0$ (eq. [\ref{p_0}]), the first two moments are  
\begin{eqnarray}
\langle\Delta\nu\rangle_0 &=& 0,
\nonumber\\
\langle(\Delta\nu)^2\rangle_0 &=& \nu^2\frac{k\te}{\me c^2},
\label{p_0_moments}
\end{eqnarray}
and the normalization is
\begin{equation}
\int P_0\,d\nu^\prime=1+\frac{3}{2}\frac{k\te}{\me c^2}.
\label{p_0_norm} 
\end{equation}

The corresponding relations for the Kompaneets equation kernel
(\ref{p_k}) are
\begin{eqnarray}
\langle\Delta\nu\rangle_k &=& \nu\left(4\frac{k\te}{\me c^2}
-\frac{h\nu}{\me c^2}\right),
\nonumber\\
\langle(\Delta\nu)^2\rangle_k &=& 2\nu^2\frac{k\te}{\me c^2},
\label{p_k_moments}
\end{eqnarray}
and 
\begin{equation}
\int
P_K\,d\nu^\prime=1+\left(\frac{5}{2}\frac{k\te}{\me
c^2}-\frac{h\nu}{\me c^2}
\right).
\label{p_k_norm} 
\end{equation}
The terms in parentheses in equation (\ref{p_k_norm}) describe the
inaccuracy of 
the $P_K$ kernel. In particular, the Klein-Nishina reduction of the
integral cross section is not covered by this approximation. The
$P_K$ kernel also neglects altogether the frequency diffusion of photons due
to Compton recoil (as a result of the recoil-induced frequency shift
depending on the scattering angle). The term of order $(h\nu/\me c^2)^2$ in
the expansion series for $\langle (\Delta\nu)^2\rangle$, which is present in
equation (\ref{p_moments}) and absent from equation (\ref{p_k_moments}), is
responsible for this. The Kompaneets equation (\ref{komp}), which is a
Fokker-Planck equation with its coefficients governed by the moments
(\ref{p_k_moments}), does not include the corresponding dispersion term, the
importance of which for the case $h\nu\gg k\te$ was pointed out in
\citep{rosetal78,illetal79}.

Having calculated the normalization (\ref{p_norm}) for the $P$ kernel
and knowing the exact result for the total scattering cross section (eq.
[\ref{sigma}]), we can try to crudely take into account the terms
of order $\eta^2$ for $P$, not calculating them. To this end, we have to
renormalize the kernel as 
\begin{equation}
P^\prime=\left[1+\frac{53}{10}\left(\frac{k\te}{\me c^2}\right)^2
+\frac{19}{5}\frac{h\nu}{\me c^2}\frac{k\te}{\me c^2}
+\frac{41}{20}\left(\frac{h\nu}{\me c^2}
\right)^2\right]P,
\label{p_rn}
\end{equation}
where $P$ is given by equation (\ref{p_set}). As a result, the
normalization of the modified kernel $P^\prime$ is precisely the
bracketed expression in equation (\ref{sigma}).

In the case $h\nu\ll k\te$, the kernel given by equation
(\ref{pt}) can be renormalized similarly:
\begin{equation}
P_T^\prime=\left[1+\frac{53}{10}\left(\frac{k\te}{\me c^2}
\right)^2\right]P_T, 
\label{pt_rn}
\end{equation}

Finally, we may introduce the renormalized kernels $P_0^\prime$ and
$P_K^\prime$:
\begin{equation}
P_0^\prime=\left(1-\frac{3}{2}\frac{k\te}{\me c^2}\right)P_0,
\label{p_0_rn}
\end{equation} 
\begin{equation}
P_K^\prime=\left(1-\frac{5}{2}\frac{k\te}{\me c^2}-\frac{h\nu}{\me c^2}
\right)P_K.
\label{p_k_rn}
\end{equation}
The assumed normalizations of $P_0^\prime$ and $P_K^\prime$ are 1 and
$1-2h\nu/\me c^2$, respectively.

We should stress that the renormalized kernels are still
approximations of the same order of uncertainty as the original
kernels, but these kernels turn out to be more accurate than the original
ones, as follows from a comparison with results of numerical
calculations, which we present in the next paragraph. Note also that
the moments are not affected by the renormalization procedure.

\subsubsection{Comparison of the Analytical Approximations for the
Kernel with Numerical Results} 

\paragraph{The case of $\bf h\bg{\nu}\ll k T_{\rm\bf e}\ll m_{\rm\bf e} c^2$.}

In this case, the profile of a Compton-scattered monochromatic line forms
through the Doppler mechanism alone. The exact kernel can be computed
numerically, using equation (\ref{p_f}) of \S 6 or by means of Monte Carlo
simulations. We employ both methods in our analysis. Figures~8 and 9
compare, for two values of the electron temperature ($k\te=10$~keV and
25~keV), the accurate spectra with the corresponding analytical dependences
as calculated in different approximations for the kernel: $P_T$, $P_0$, and
$P_K$ (eqs. [\ref{pt},\ref{p_0},\ref{p_k}]).

As expected, the asymmetry of the line (domination of the right wing over
the left) increases as the temperature grows. One can see that the zero
approximation \citep{hummih67}, $P_0$, which is symmetric in frequency
shift, matches the line profile poorly. Therefore, we recommend its usage be 
restricted to the range $k\te\lesssim 500$~eV, where $P_0$ is accurate to
better than 98\%, except in the far wings of the line. The latter can be
roughly defined as the regions $|\nu^\prime-\nu|/(\nu^\prime+\nu)\gtrsim 1/4$.

\begin{figure*}[tb]
\epsfxsize=16.5cm
\epsffile[-110 190 685 700]{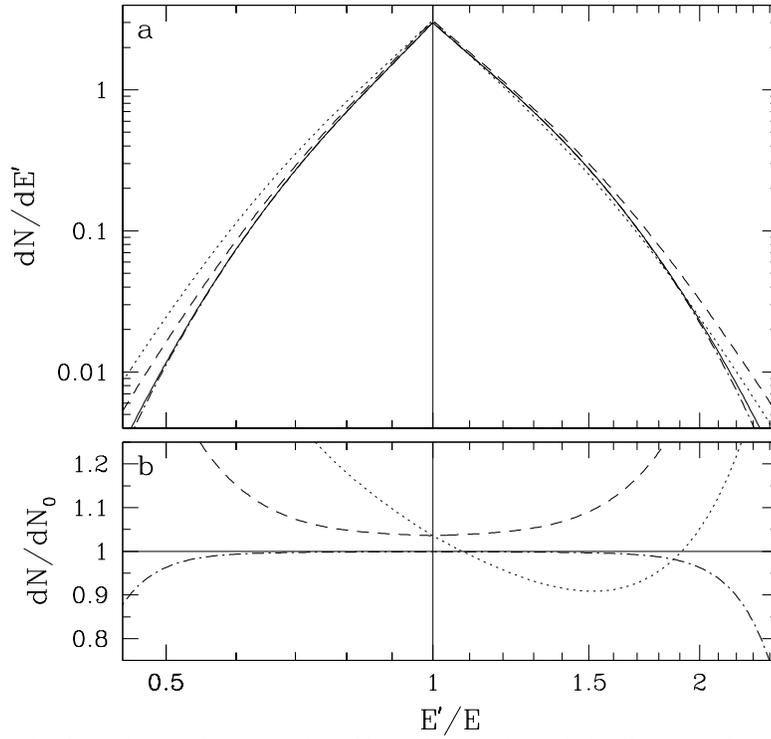}
\caption{
(a) Spectrum that forms through single scattering of isotropic
monochromatic low-frequency ($h\nu\ll k\te$) radiation on weakly
relativistic thermal electrons, $k\te=10$~keV. The solid line
shows the result of an accurate numerical calculation from
eq. (\ref{p_f}). Also shown are the approximations given by:
eq. (\ref{p_0}), the zero-order kernel $P_0$ ({\sl dotted line});
eq. (\ref{p_k}), the Kompaneets equation kernel $P_K$ ({\sl dashed
line}); and eq. (\ref{pt}), the mildly relativistic kernel $P_T$ ({\sl
dash-dotted line}). (b) Ratio of the approximate spectra shown
in (a) to the accurate spectrum.
}
\end{figure*}

\begin{figure*}[tb]
\epsfxsize=16.5cm
\epsffile[-110 190 685 700]{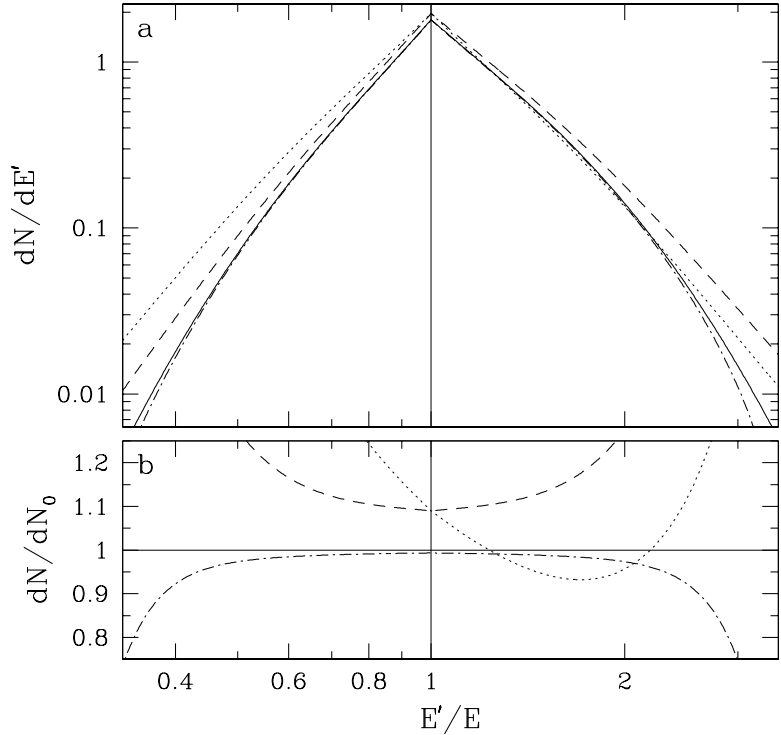}
\caption{
Same as Fig.~8, but for a higher temperature, $k\te=25$~keV.
}
\end{figure*}

\begin{figure*}[tb]
\epsfxsize=16.5cm
\epsffile[-110 190 685 700]{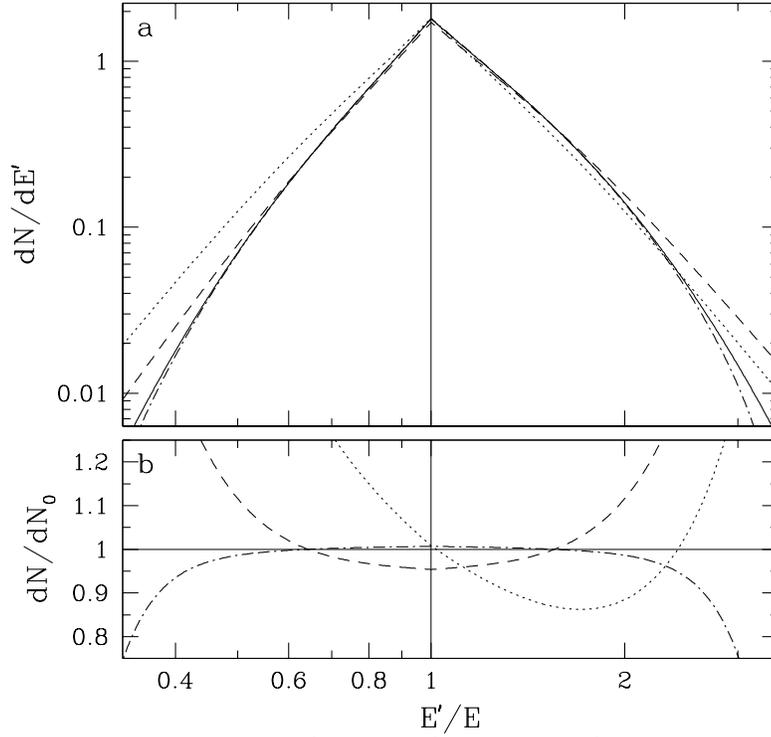}
\caption{
Same as Fig.~9, but the renormalized kernels
$P_0^\prime$ (eq. [\ref{p_0_rn}]), $P_K^\prime$ (eq. [\ref{p_k_rn}]),
and $P_T^\prime$ (eq. [\ref{pt_rn}]), are used. One can see that the agreement
between the approximations and the exact kernel is better than in Fig.~9.
}
\end{figure*}

\begin{figure*}[tb]
\epsfxsize=16.5cm
\epsffile[-110 190 685 700]{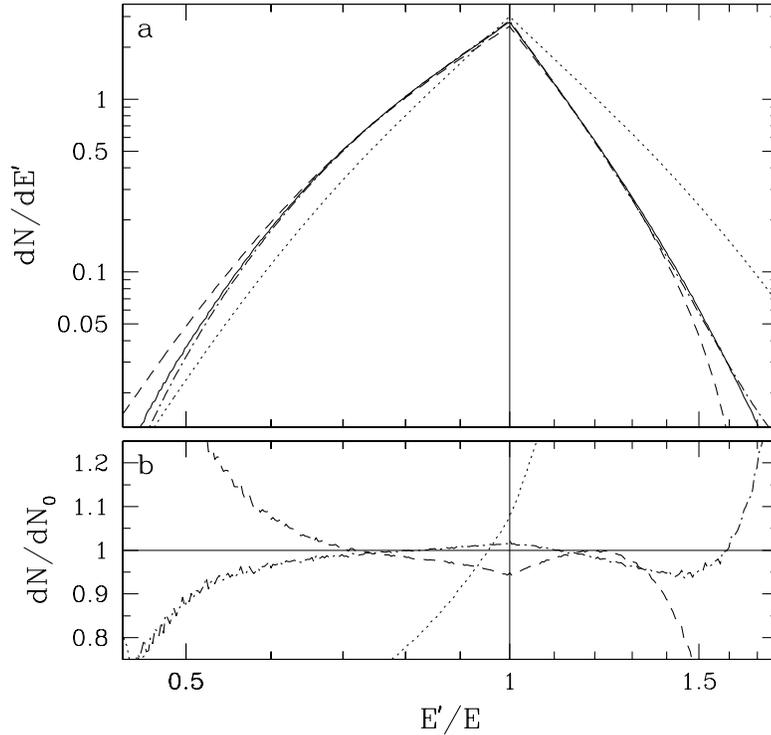}
\caption{
(a) Spectrum that forms through single scattering of isotropic
monochromatic radiation of high energy, $h\nu=50$~keV, on weakly
relativistic thermal electrons, $k\te=10$~keV. The result of a
Monte Carlo simulation ({\sl solid line}) is compared with the
different approximations for the kernel: $P_0^\prime$
(eq. [\ref{p_0_rn}], {\sl dotted line}), $P_K^\prime$
(eq. [\ref{p_k_rn}], {\sl dashed line}), and $P^\prime$
(eq. [\ref{p_rn}], {\sl dash-dotted line}). (b) Ratio 
of the approximate spectra shown in (a) to the accurate (Monte Carlo) spectrum.
}
\end{figure*}

\begin{figure*}[tb]
\epsfxsize=16.5cm
\epsffile[-110 190 685 700]{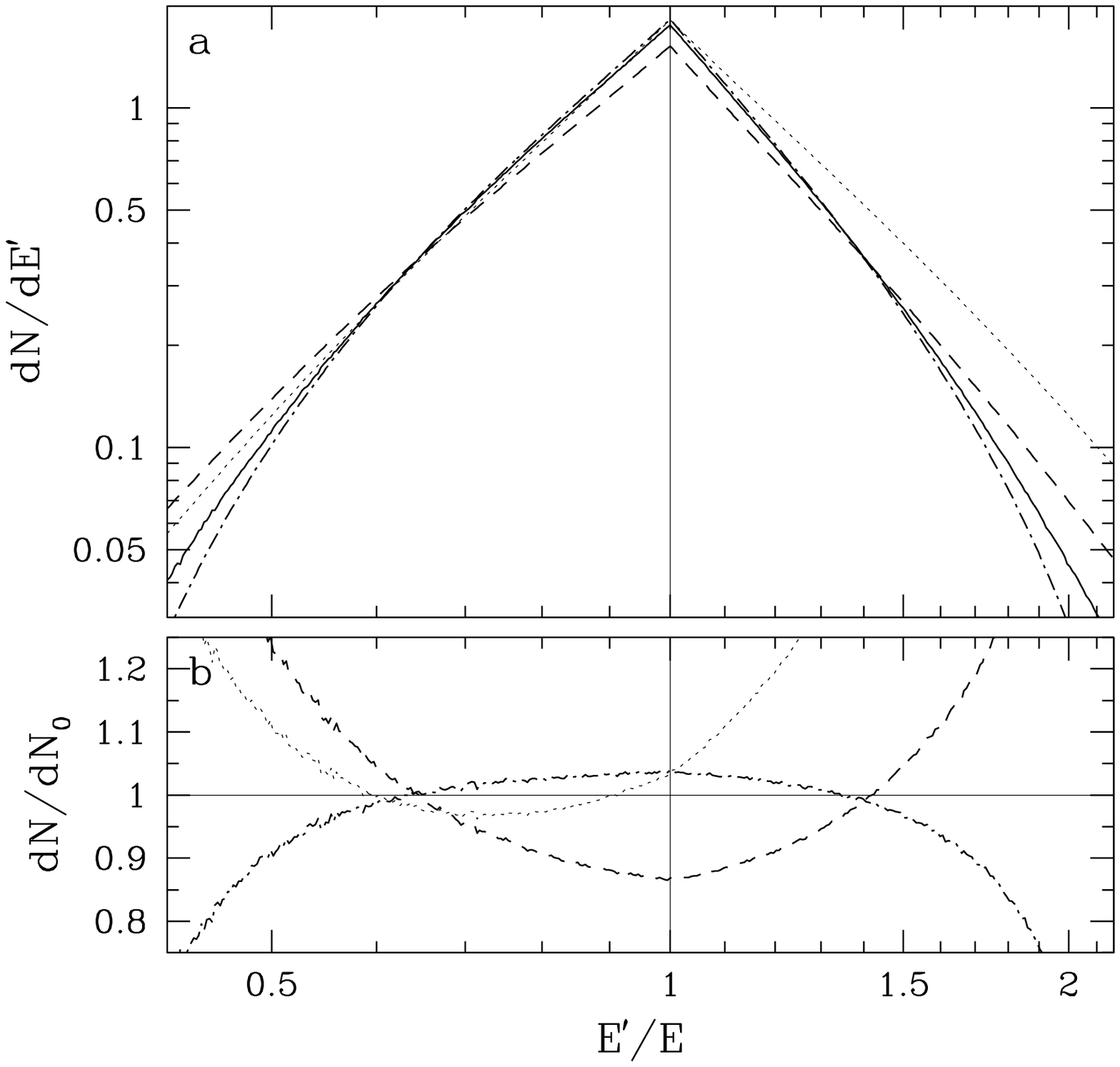}
\caption{
Same as Fig.~11 ($h\nu=50$~keV), but for a higher
electron temperature, $k\te=25$~keV.
}
\end{figure*}

The Kompaneets equation kernel, $P_K$, which was derived for the considered
case ($h\nu\ll k\te$) by \cite{sunyaev80}, works well when $k\te\lesssim
5$~keV. At higher temperatures, it becomes important to take into account the
relativistic correction terms, $(1+\sqrt{2}\delta\eta^{1/2}) p_t$, in
equation (\ref{pt}), i.e., to use our most accurate approximation, $P_T$. At
$k\te=25$~keV, the \citep{sunyaev80} kernel overestimates the flux at the
line peak by 10\% (see Fig.~9). The flux turns out to be even more
overestimated in the wings. The $P_T$ kernel is accurate to better than
98\% at this temperature.

At $k\te\sim 50$~keV, the electrons are already strongly
relativistic: $\langle (v/c)^2\rangle^{1/2}\sim \eta^{1/2}\sim 0.5$, 
and the $\eta^{1/2}$ series in equation (\ref{pt}) converges poorly,
particularly in the wings of the line. At higher temperatures, one must use
the exact kernel (which can be calculated from eq. [\ref{p_f}]) and treat
the problem numerically.

Figure~10 is the same as Figure~9, but presents the results for the modified
kernels, $P_T^\prime$, $P_0^\prime$, and $P_K^\prime$ (eqs.
[\ref{pt_rn}]--[\ref{p_k_rn}]). It is seen that the renormalization has
indeed appreciably improved the accuracy of the approximations.

\paragraph{Inclusion of quantum effects: $\bf h\bg{\nu}\neq 0$, $\bf k
T_{\rm\bf e}\ll m_{\rm\bf e} c^2$.}

The assumption $h\nu(h\nu/\me c^2)\ll k\te$, which we used when deriving
equation (\ref{p_set}), means that the recoil-induced downward shift in the
photon frequency must be small compared to the Doppler line broadening.
Using Monte Carlo simulations, we have established that $P$ and $P_K$ (eqs.
[\ref{p_set}] and [\ref{p_k}]) remain good approximations up to $h\nu(h\nu/\me
c^2)\sim k\te$, i.e., when the Doppler and recoil effects become comparable.
Naturally, the zero-order kernel, $P_0$, is a very poor approximation in this
case, because it totally neglects Compton recoil. This is demonstrated
by Figures~11 and 12, which compare accurate (Monte Carlo)
single-scattering spectra with the corresponding profiles computed from the
renormalized kernels $P^\prime$, $P_0^\prime$, and $P_K^\prime$
(eqs. [\ref{p_rn}], [\ref{p_0_rn}], and [\ref{p_k_rn}]).

The results of our simulations imply that the temperature limits quoted above
for the case $h\nu\ll k\te$ remain valid upon inclusion of quantum effects.
We finally conclude that the $P$ kernel can be used in the following range
of parameter values: $h\nu(h\nu/\me c^2)\lesssim k\te\lesssim 25$~keV,
$h\nu\lesssim 50$~keV. The corresponding limits for the $P_K$ kernel are
$h\nu(h\nu/\me c^2)\lesssim k\te\lesssim 5$~keV,
$h\nu\lesssim 50$~keV.

At $h\nu(h\nu/\me c^2)\gtrsim k\te$, recoil becomes more important than the
Doppler effect. The single-scattered profile becomes double-peaked
\citep[see, e.g., the results of Monte Carlo simulations of][]{pozetal83}. This
case requires a special analytical treatment that is beyond the scope of
this paper. Here we may only point out the principle mathematical diffuculty
that makes it impossible to use our current method for calculating the
$P(\nu\rightarrow\nu^\prime)$ kernel in this limit; the exponential
entering the expression for $K$ (eq. [\ref{k_set}]) cannot be expanded in
powers of $\eta^{1/2}$, as was done for the case $h\nu(h\nu/\me c^2)\lesssim
k\te$.

\begin{figure*}[tb]
\epsfxsize=16.5cm
\epsffile[-110 190 685 700]{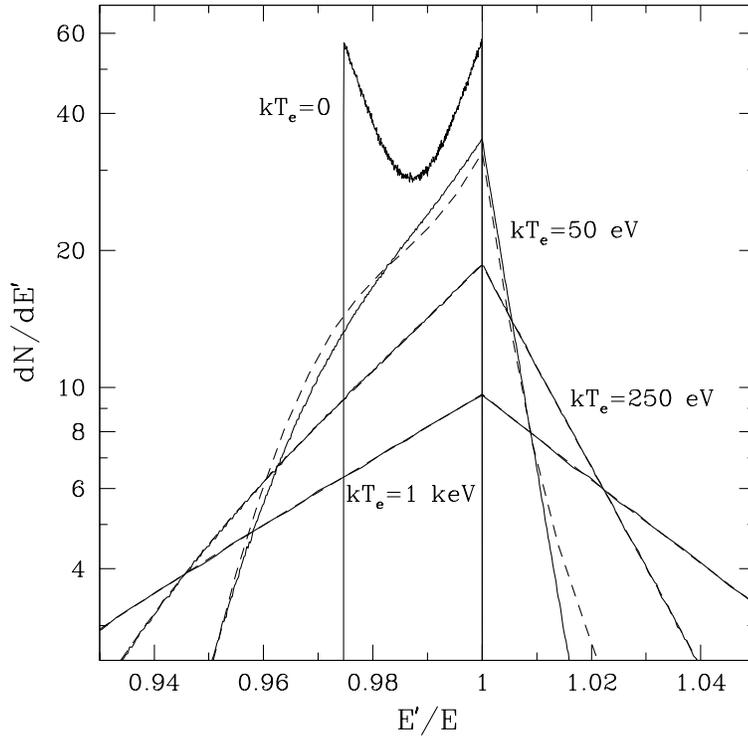}
\caption{
Spectra resulting from the single scattering of isotropic
monochromatic radiation of energy $h\nu=6.7$~keV on low-temperature
($h\nu> 4 k\te$) thermal electrons, for different values of
$k\te$. In this case, the Compton-recoil shift is larger than the Doppler
shift. The results of Monte Carlo simulations ({\sl solid lines}) are
compared with the results of the calculation by the approximate eq.
(\ref{p_rn}) for the mildly relativistic kernel $P^\prime$ ({\sl
dashed lines}). For the case $k\te=0$ (cold electrons), only the
Monte Carlo result ({\sl double-peaked profile}) is shown, since our
approximation for the kernel is not valid in this case.
}
\end{figure*}

\begin{figure*}[tb]
\epsfxsize=16.5cm
\epsffile[-110 190 685 700]{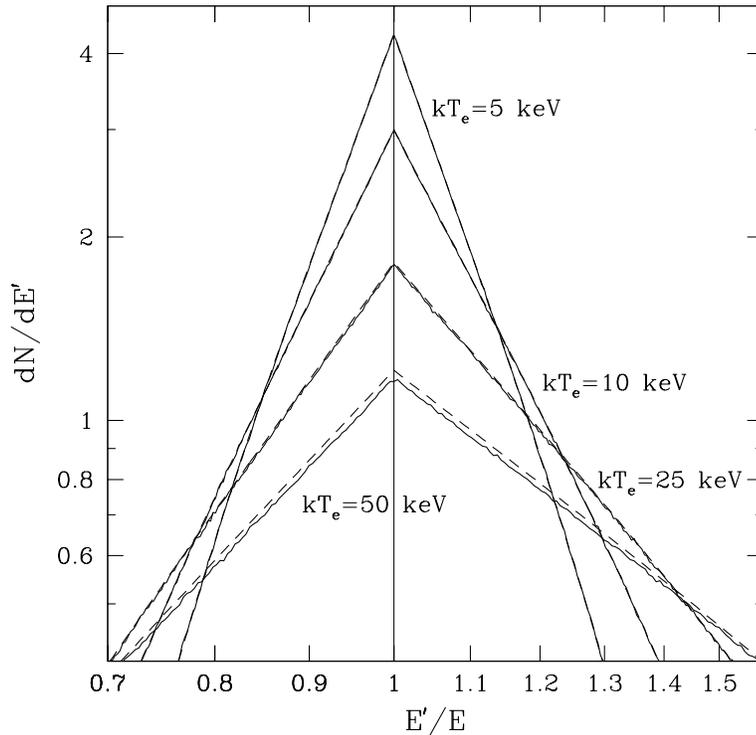}
\caption{
Same as Fig.~13 ($h\nu=6.7$~keV), but for
high-temperature electrons, $h\nu< 4 k\te$. In this case, the Doppler
shift is larger than the recoil shift.
}
\end{figure*}

\begin{figure*}[tb]
\epsfxsize=16.5cm
\epsffile[-110 190 685 700]{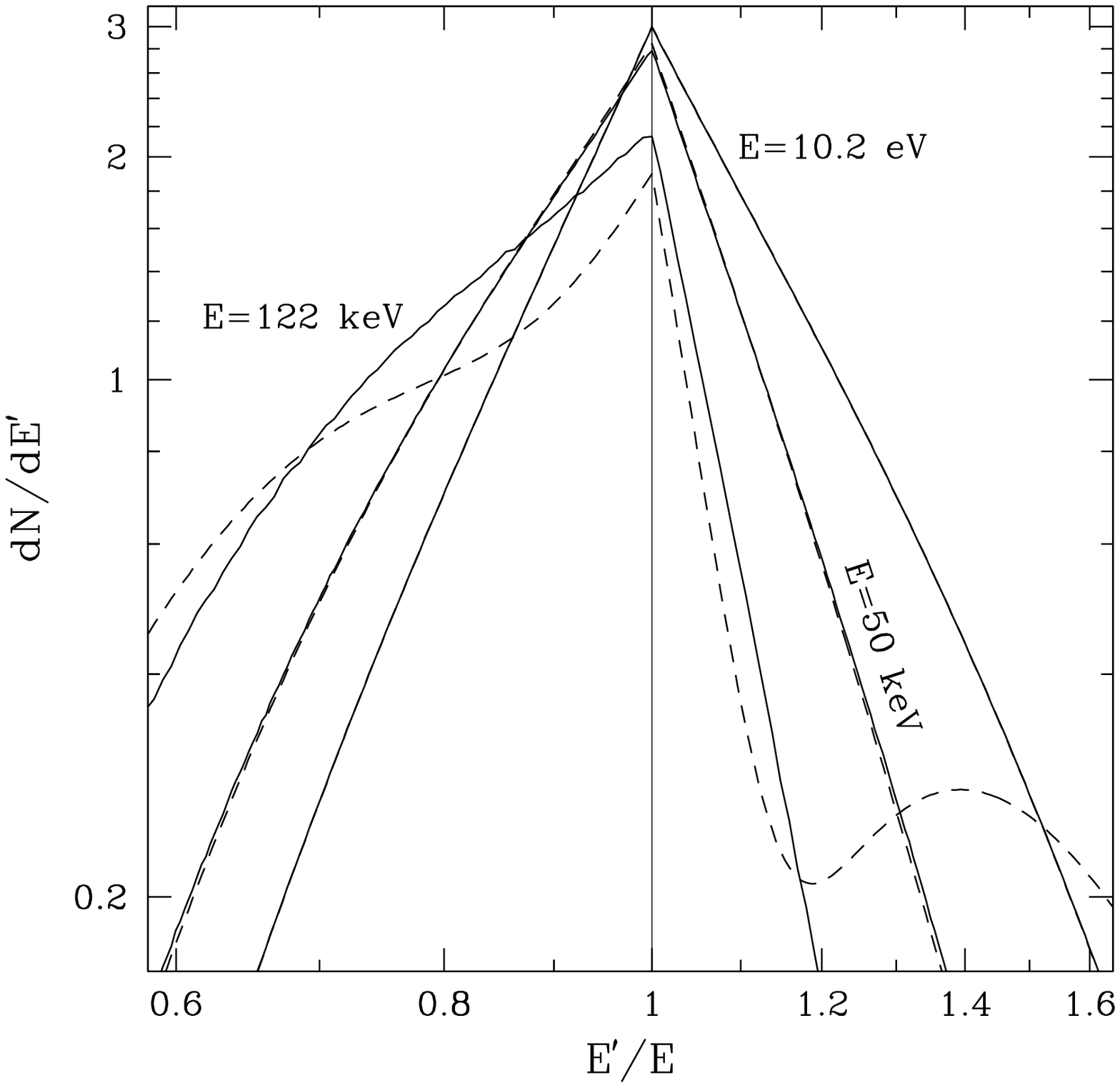}
\caption{
Spectra resulting from the single scattering of isotropic
monochromatic radiation on weakly relativistic electrons,
$k\te=10$~keV, for different photon energies. The results of
Monte Carlo simulations ({\sl solid lines}) are compared with the results of
the calculation by the approximate eq. (\ref{p_rn}) for the mildly
relativistic kernel $P^\prime$ ({\sl dashed lines}). One can observe how the
effect of Compton recoil on the spectrum increases as the photon energy
becomes higher. The case of the 122~keV nuclear line produced by $^{57}$Co
is beyond the scope of our analytical approximation for the isotropic kernel.
}
\end{figure*}

Numerous examples of spectra forming by single Compton scattering, which may
well be encountered in various astrophysical situations, and which are well
approximated by our analytical kernel equation (\ref{p_set}), are presented in
Figures~13--15.

\subsubsection{Properties of the Single-Scattering Profile}

The line profile forming by single scattering of isotropic radiation
on thermal electrons (see Figs.~8--15), which is approximated by
equation (\ref{p_set}), is unique in its properties. It therefore appears
interesting to examine its characteristic features in some detail (following 
Pozdnyakov et al. 1983).
 
First, let us compare the line profile as calculated from equation
(\ref{p_set}) with the usual Gaussian profile that, e.g., may result from
Doppler broadening of an emission line in the presence of 
thermal or turbulent motions of ions. To simplify this comparison, let us
assume that $h\nu\ll k\te$. For a given plasma temperature, it is  
natural to adopt $\Delta\nu_D=\nu(2\eta)^{1/2}$ for the Doppler
shift\footnote{Note that the width used, $\Delta\nu_D$, is
$(M/\me)^{1/2}=43(M/m_{\rm p})^{1/2}$ times the thermal width of
lines of an ion of mass $M$.}, i.e. $N\sim
\exp{[-(\nu^\prime-\nu)^2/\Delta\nu_D^2]}$. The mean (rms) 
frequency shift, $\sqrt{\langle(\Delta\nu)^2\rangle}$,
is $\nu\eta^{1/2}$ for the Doppler profile. The corresponding value for
the Compton-scattered line is $\nu[2\eta(1+23.5\eta)]^{1/2}$ (as
results from the value of the second moment given by
eq. [\ref{p_moments}]). The width of the single-scattering profile at
half-maximum (FWHM) is approximately equal to
$2\nu[\ln{2}\eta]^{1/2}=1.66\nu\eta^{1/2}$. The corresponding value for 
the Doppler profile is $2\nu[2\ln{2}\eta]^{1/2}$, i.e., $\sqrt{2}$ times
more, which is opposite to the situation with the rms shift. Thus, in
the case of the line forming by Compton scattering, 
relatively few photons appear in the upper part of the profile (above
half-maximum), and an accordingly large fraction of the scattered
radiation emerges in the wings of the line. It is also worth noting
that the Doppler profile is symmetric, while the profile due to Compton
scattering is not.

Now let us consider the peak of the single-scattering profile, a
detail which makes it so peculiar. In the vicinity of the maximum
($|\nu^\prime-\nu|\ll\nu\eta^{1/2}$), the spectrum is well
approximated by the following expression, which results from equation
(\ref{p_set}):
\begin{eqnarray}
P(\nu\rightarrow\nu^\prime)_{+,-}=\nu^{-1}\frac{11}{20}\sqrt{\frac{2}{\pi}}
\eta^{-1/2}
\left\{\left[1+\left(-\frac{1091}{616}-\frac{23}{154}\left(\frac{h\nu}
{k\te}\right)^2\right)\eta\right]
\right.
\nonumber\\
\left.
+\frac{\nu^\prime/\nu-1}{\eta^{1/2}}\left[\mp\frac{15}{22}\sqrt{\frac{\pi}{2}}
+\frac{1}{2}\left(1-\frac{h\nu}{k\te}\right)\eta^{1/2}+...\right]
+...\right\}, 
\label{peak}
\end{eqnarray}
where the indices plu and minus signs correspond to the right and left
wings, respectively.

We see that the spectrum has a cusp at $\nu^\prime=\nu$ (a break in
the derivative occurs there). Near the cusp, on its both sides, the
spectrum can be approximated as a power law, the slopes in the right
and left wings [the coefficient at $(\nu^\prime/\nu-1)$ in
eq. (\ref{peak})] being 
significantly different: 
\begin{eqnarray}
\alpha_{+} &=& -\left[\frac{d\ln{P}}{d\ln{(\nu^\prime/\nu)}}\right]_{\nu^
\prime=\nu+0}=
\frac{15}{22}\sqrt{\frac{\pi}{2}}\eta^{-1/2}
-\frac{1}{2}+\frac{1}{2}\frac{h\nu}{k\te},
\nonumber\\
\alpha_{-} &=& ~~\left[\frac{d\ln{P}}{d\ln{(\nu^\prime/\nu)}}\right]_{\nu^
\prime=\nu-0}=\frac{15}{22}\sqrt{\frac{\pi}{2}}\eta^{-1/2}
+\frac{1}{2}-\frac{1}{2}\frac{h\nu}{k\te},~~~~~~
\alpha_{-}-\alpha_{+}=1-\frac{h\nu}{k\te}
\label{alphas}
\end{eqnarray}
(in Pozdnyakov et al. 1983, $\alpha_{-}-\alpha_{+}=3$ when $h\nu=0$,
because they considered the energy spectrum that is the
product of $\nu^\prime/\nu$ and the photon spectrum considered here).

It is interesting that when $h\nu=k\te$, the line profile in the
vicinity of the cusp is symmetric (in logarithmic coordinates) about
$\nu^\prime=\nu$ 
($\alpha_{+}=\alpha_{-}$). Let us give a few further examples. If $h\nu=0$ and
$\eta=$~0.001, 0.01, and 0.1, the slopes run:
$\alpha_{+,-}=27.0\mp0.5$, $8.5\mp0.5$, and $2.7\mp0.5$. For
$h\nu/\me c^2=0.1$ and $\eta=0.02$ (the case
of Figure~11): $\alpha_{+}=8.0$, $\alpha_{-}=4.1$, and
$\alpha_{-}-\alpha_{+}=-3.9$ (for $h\nu=0$ and $\eta=0.02$, we would
have $\alpha_{+}=5.5$, $\alpha_{-}=6.6$, and $\alpha_{-}-\alpha_{+}=1.0$).

The asymmetry of the single-scattering profile can be demonstrated by
comparing the fractions of the scattered radiation contained in the right
($\nu^\prime>\nu$) and left ($\nu^\prime<\nu$) wings of the
line. Using the renormalized kernel given by equations (\ref{p_rn})
and (\ref{p_set}), we find in terms of number of photons ($N=\int
P^\prime\,d\nu^\prime$) and total energy ($W=\int h\nu^\prime
P^\prime\,d\nu^\prime$), 
\begin{eqnarray}
N_{+,-}=\frac{1}{2}\pm
\eta^{1/2}\sqrt{\frac{2}{\pi}}\left[\frac{69}{70}-\frac{23}{70}
\frac{h\nu}{k\te}\right]-\frac{h\nu}{\me c^2}
\pm\eta^{3/2}\sqrt{\frac{2}{\pi}}\left[-\frac{1577}{1680}-\frac{4061}{1680}
\frac{h\nu}{k\te}
\right.
\nonumber\\
\left.
+\frac{43}{84}\left(\frac{h\nu}{k\te}\right)^2
+\frac{43}{1260}\left(\frac{h\nu}{k\te}\right)^3\right]+
\eta^2\left[-\frac{5}{2}\frac
{h\nu}{k\te}+\frac{13}{5}\left(\frac{h\nu}{k\te}\right)^2\right],
\label{n_mp}
\end{eqnarray}
\begin{eqnarray}
W_{+,-}=\left\{\frac{1}{2}\pm
\eta^{1/2}\sqrt{\frac{2}{\pi}}\left[\frac{23}{14}-\frac{23}{70}\frac{h\nu}
{k\te}\right]+\eta\left[2-\frac{3}{2}\frac{h\nu}{k\te}\right]
\pm\eta^{3/2}\sqrt{\frac{2}{\pi}}\left[\frac{187}{48}-\frac{521}{80}\frac
{h\nu}{k\te}
\right.\right.
\nonumber\\
\left.\left.
+\frac{43}{6}\left(\frac{h\nu}{k\te}\right)^2
+\frac{43}{1260}\left(\frac{h\nu}{k\te}\right)^3\right]+
\eta^2\left[5-\frac{57}{4}\frac{h\nu}{k\te}+\frac{47}{10}
\left(\frac{h\nu}{k\te}\right)\right]\right\}h\nu.
\label{w_mp}
\end{eqnarray}
The renormalization is important here because it enables us to obtain the
terms $O(\eta^2)$, $O(\eta h\nu/\me c^2)$, and $O((h\nu/\me c^2)^2)$
in equations (\ref{n_mp}) and (\ref{w_mp}). This procedure
is strictly correct for the following reason. The terms of even orders
in $\eta^{1/2}$ in the power series for the $P(\nu\rightarrow\nu^\prime)$
kernel (see eq. [\ref{p_set}]), i.e. $p_0$, $p_t$, $p_r$ and analogous
(unknown) terms of higher orders, are symmetric in frequency variation.
Therefore, if we know (and we indeed do) the contribution of such a term, say
$O(\eta^2)$, to the normalization of the accurate kernel, we immediately
know that the contribution of this term to both $N_{+}$ and $N_{-}$ is
equal to half this  value. $W_{+,-}$ are then determined as $\int
h(\nu+\nu^\prime-\nu)P^\prime\,d\nu^\prime= N_{+,-}h\nu+\int
h(\nu^\prime-\nu)P^\prime\,d\nu^\prime$. The last integral is accurate to 
within $\eta^2$, $\eta h\nu/\me c^2$, and $(h\nu/\me c^2)^2$
because of the presence of the small factor $\nu^\prime-\nu$.

An interesting conclusion can be drawn from both equation (\ref{w_mp}) and the
above discussion: the left wing contributes to the total accumulation
of energy by the photons exactly as much as the right wing.

\subsubsection{Single Scattering of a Step-Function Spectrum}

Although the present paper is mainly devoted to the case of single
Compton scattering of narrow spectral lines, our basic formulae --- equation
(\ref{k_set}) for the direction-dependent problem and equation (\ref{p_set})
for the isotropic problem --- describe the kernels of the corresponding
integral kinetic equations (eqs. [\ref{kinetic_k}] and
[\ref{kinetic}]) appearing 
in the general problem of Comptonization in thermal plasmas. Let us present a
simple example of using the $P(\nu\rightarrow \nu^\prime)$
kernel. Consider the scattering of an isotropic photon distribution
described by the step function, i.e., $dN_0/d\nu=1$ if $\nu\le\nu_0$ and
$dN_0/d\nu=0$ if $\nu>\nu_0$, in an optically thin (so that multiple
scatterings are not important), hot plasma. The spectrum of the scattered
photons can be found by convolving the initial frequency distribution with
the kernel:
\begin{equation}
\frac{dN(\nu)}{d\nu}=\tau\int\frac{dN_0(\nu^\prime)}{d\nu}
P(\nu^\prime\rightarrow\nu)\,d\nu^\prime,
\label{step_gen}
\end{equation}
where $\tau\ll 1$ is the Thomson optical depth of the scattering medium.

Figure~16 presents examples of spectra of the scattered component as
calculated from equation (\ref{step_gen}) using different analytical
approximations for the kernel: $P_0^\prime$ (eq. [\ref{p_0_rn}]), $P_K^\prime$
(eq. [\ref{p_k_rn}]), and $P^\prime$ (eq. [\ref{p_rn}]). The radiation
is assumed to be low-frequency ($h\nu_0\ll k\te$). In one
case, $k\te=1$~keV, a spectrum forming in the case of 
non-negligible photon energy ($h\nu_0=7.1$~keV) is also shown. The
integral in equation (\ref{step_gen}) was performed numerically. Note that
the actual observable spectrum will be the sum of the unscattered and
scattered components, and will depend on $\tau$, i.e., $(dN/d\nu)_{\rm
total}=(1-\tau)dN_0/d\nu+\tau dN/d\nu$.

\begin{figure*}[tb]
\epsfxsize=16.5cm
\epsffile[-110 190 685 700]{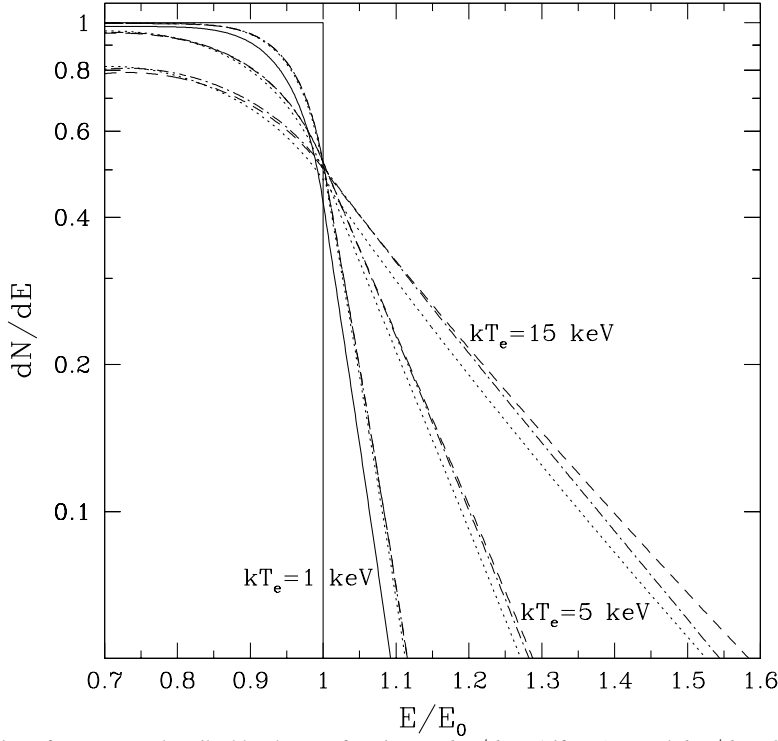}
\caption{
Single scattering of a spectrum described by the step
function --- $dN_0/dE=1$ if $E\le E_0$ and $dN_0/dE=0$ if
$E>E_0$ --- on mildly relativistic thermal electrons,
considered for various electron temperatures, assuming that the
photon energy is negligible ($E_0\ll k\te$). The spectra resulting 
from the convolution (eq. [\ref{step_gen}]) of the step function with
the following approximations for the isotropic kernel are shown:
the zero-order kernel $P_0^\prime$ (eq. [\ref{p_0_rn}], {\sl dotted
lines}), the Kompaneets equation kernel $P_K^\prime$
(eq. [\ref{p_k_rn}], {\sl dashed lines}), and the mildly
relativistic kernel $P^\prime$ (eq. [\ref{p_rn}], {\sl dash-dotted
lines}). For the case of $k\te=1$~keV, also shown is the spectrum
(corresponding to the $P^\prime$ kernel) for a nonnegligible photon energy,
$E_0=7.1$~keV ({\sl solid line}). Note that the actual observable
spectrum will be the sum of the nonscattered and scattered
components, i.e., $(dN/dE)_{\rm total}=(1-\tau)dN_0/dE+\tau dN/dE$,
where $\tau\ll 1$ is the Thomson optical depth of the scattering
medium.
}
\end{figure*}

The spectra shown in Figure~16 allow analytical description. We
present here only the result of the convolution of the step function with
the Kompaneets equation kernel, $P_K$:
\begin{mathletters}
\begin{equation}
\frac{dN}{d\nu}=\left\{\begin{array}{ll}
  (f_0+f_1)\tau, & \nu > \nu_0\\
(1-f_0+f_1)\tau, & \nu\le\nu_0,
\end{array}\right.\
\label{step}
\end{equation}
where
\begin{eqnarray}
f_0=\frac{1}{\sqrt{\pi}}\left[\left(-\frac{6}{5}\delta_0
-\frac{13}{15}\delta_0^3-\frac{4}{15}\delta_0^5\right)F+\left(1+3\delta_0^2
+2\delta_0^4+\frac{8}{15}\delta_0^6\right)G\right],
\nonumber\\
f_1=\sqrt{\frac{2}{\pi}}\left[\left(\frac{23}{70}-\frac{27}{35}\delta_0^2
-\frac{24}{35}\delta_0^4-\frac{8}{35}\delta_0^6\right)F+\left(2\delta_0^3
+\frac{8}{5}\delta_0^5+\frac{16}{35}\delta_0^7\right)G\right]\eta^{1/2}
\left(1-\frac{h\nu_0}{k\te}\right),
\nonumber\\
F=\exp{(-\delta_0^2)},\,\,\,G=\int_{\delta_0}^{\infty}\exp{(-t^2)}\,dt=0.5
\pi^{1/2}Erfc(\delta_0),\,\,\,\delta_0=(2\eta)^{-1/2}\frac{|\nu-\nu_0|}
{\nu_0+\nu}.
\end{eqnarray}
\label{step_set}
\end{mathletters}

The main term in equation (\ref{step}) ---  $f_0\tau$ (if
$\nu>\nu_0$) or $(1-f_0)\tau$ (if $\nu\le\nu_0$) ---  results from the
zero-order kernel, $P_0$ \citep{hummih67}. 

The spectrum of the scattered component at frequencies $\nu>\nu_0$ has
a quasi-power-law shape, with a slope that is approximately equal to
the slope of the right wing of the kernel itself (see eq. [\ref{alphas}]):
\begin{equation}
\alpha=\frac{11}{10}\sqrt{\frac{2}{\pi}}\eta^{-1/2}-\frac{253}{175\pi}\left(
1-\frac{h\nu_0}{k\te}\right).
\label{step_slope}
\end{equation}
As is the case with the kernel itself, the first-order temperature
correction causes the spectrum of the scattered component to be
flatter than results from the zero-order approximation. Compton recoil
has an opposite effect on the slope and may cause a significant
steepening of the spectrum when $h\nu_0\gg k\te$. It is clear from
Figure~16 and equation (\ref{step_slope}) that in the low-frequency
case, it is possible to determine the temperature of the scattering
plasma by measuring the slope of the spectrum of the scattered component.

Let us also give an expression for the total number of photons that
have been redistributed from the frequency region $\nu\le\nu_0$ into the
region $\nu>\nu_0$:
\begin{equation}
N=\left[\frac{23}{35}\sqrt{\frac{2}{\pi}}+\left(1-\frac{h\nu_0}{\me c^2}\right)
\eta^{1/2}\right]\nu_0\eta^{1/2}\tau. 
\end{equation} 

We can also imagine a situation in which a spectrum described by the
left-side step function: $dN_0/d\nu=1$ if $\nu\ge\nu_0$ and $dN_0/d\nu=0$ if
$\nu<\nu_0$, is being scattered. In this case, we arrive at a formula that
is similar to equation (\ref{step_set}):
\begin{equation}
\frac{dN}{d\nu}=\left\{\begin{array}{ll}
  (f_0-f_1)\tau, & \nu < \nu_0\\
(1-f_0-f_1)\tau, & \nu\ge\nu_0.
\end{array}\right.\
\label{step_left}
\end{equation}\
We see that here the asymmetry of the kernel has the opposite
effect on the scattered spectrum as compared with the previous
situation, in which the right-side step function was considered
(compare the signs of the term $f_1$ in eqs. [\ref{step}] and
[\ref{step_left}]), namely, the temperature correction causes the spectrum to
steepen, while the recoil correction makes the slope flatter.

\subsection{Fokker-Planck Expansion of the Integral Kinetic Equation}

We can carry out a Fokker-Planck expansion (eq. [\ref{fokker}]) of the
kinetic equation (\ref{kinetic}) with the $P(\nu\rightarrow\nu^\prime)$
kernel derived in this paper (eq. [\ref{p_set}]). The coefficients in
equation (\ref{fokker}) depend on the moments of the kernel, which are given
by equation (\ref{p_moments}). As a result, we obtain the generalized (for
the mildly relativistic case) Kompaneets equation: 
\begin{eqnarray}
\frac{\partial n(\nu)}{\partial\tau}=\frac{h}{\me c^2}\frac{1}{\nu^2}
\frac{\partial}{\partial \nu}\nu^4\left\{n(1+n)+\frac{k\te}{h}
\frac{\partial n}{\partial\nu}+\frac{7}{10}\frac{h\nu^2}{\me c^2}
\frac{\partial n}{\partial\nu}+\frac{k\te}{\me c^2}\left[\frac{5}{2}\left(
n(1+n)+\frac{k\te}{h}\frac{\partial n}{\partial\nu}\right)
\right.\right.
\nonumber\\
\left.\left.
+\frac{21}{5}\nu \frac{\partial}{\partial\nu}\left(n(1+n)+\frac{k\te}{h}
\frac{\partial n} {\partial\nu}\right)
+\frac{7}{10}\nu^2\left(-2\left(\frac{\partial n}{\partial\nu}\right)^2+2(1+2n)
\frac{\partial^2 n}{\partial\nu^2}+\frac{k\te}{h}\frac{\partial^3 n}{\partial
\nu^3}\right)\right]\right\}.
\label{komp_gen}
\end{eqnarray}

This equation was earlier obtained, in a different way, by \cite{itoetal98}
and \cite{chalas98}. Substituting the Planckian distribution
that corresponds to the temperature of the electrons,
$n=(e^{h\nu/k\te}-1)^{-1}$, into equation (\ref{komp_gen})
yields $\partial n(\nu)/\partial\tau=0$. This test confirms once more that
equation (\ref{p_set}) for the kernel consistently takes into account
necessary corrections. One can also make sure that equation
(\ref{komp_gen}) without the terms responsible for induced scattering does
not modify a Wien spectrum, $n=e^{-h\nu/k\te}$. The fact that equation
(\ref{komp_gen}) conserves the total number of photons follows directly from
its divergent form. Thus, as expected, the basic properties of the
Kompaneets equation are retained in equation (\ref{komp_gen}).

Equation (\ref{komp_gen}) can be simplified in the case $h\nu\ll
k\te$. Omitting the terms related to quantum effects and induced
scattering, we find
\begin{equation}
\frac{\partial n(\nu)}{\partial\tau}=\frac{k\te}{\me c^2}\frac{1}
{\nu^2}\frac{\partial}{\partial \nu}\nu^4\left[\frac{\partial n}{\partial\nu}
+\frac{k\te}{\me c^2}\left(\frac{5}{2}\frac{\partial n}{\partial\nu}
+\frac{21}{5}\nu\frac{\partial^2 n}{\partial\nu^2}+\frac{7}{10}\nu^2
\frac{\partial^3 n}{\partial\nu^3}\right)\right].
\label{komp_gen_a}
\end{equation}
This equation allows one to derive the first-order relativistic
corrections to the effect of distortion of the CMB spectrum in clusters of
galaxies with hot gas \citep[][see another way of finding these
analytical corrections in Sazonov \& Sunyaev 1998]{itoetal98,chalas98}.

In another limit, $h\nu(h\nu/\me c^2)\gg k\te$, when the Doppler effect is
not important, one obtains (ignoring induced-scattering terms)
\begin{equation}
\frac{\partial n(\nu)}{\partial\tau}=\frac{h}{\me c^2}\frac{1}{\nu^2}
\frac{\partial}{\partial
\nu}\nu^4\left(n+\frac{7}{10}\frac{h\nu^2}{\me c^2} 
\frac{\partial n}{\partial\nu}\right).
\label{komp_gen_b}
\end{equation}
The second parenthesized term in this equation, which describes frequency
diffusion of photons, was added to the Kompaneets equation by
\cite{rosetal78} and \cite{illetal79}. This correction becomes 
especially important when studying the scattering of hard radiation 
on cold electrons in an optically thick medium. Such a situation takes
place, for example, during a supernova explosion; an analytical
solution to the corresponding diffusion problem was derived and
employed in a calculation of the evolution of the X-ray spectrum of
Supernova 1987A by \cite{gresun87}.

In the intermediate case of $h\nu\gg k\te$, $h\nu(h\nu/\me c^2)\ll
k\te$, the dispersion term $(k\te/\me c^2)\nu^{-2}\partial/
\partial\nu(\nu^4\partial n/\partial\nu)$ of the Kompaneets
equation, which describes the diffusion of the photons due to the
Doppler effect, must be added to equation (\ref{komp_gen_b}).

\subsection{Kinetic Equation for Problems with a Decisive Role
of Induced Compton Scattering}

If the conditions $n\gg 1$ and $n\gg k\te/h\nu$ are both satisfied,
then equation (\ref{komp_gen}) will simplify to (only induced-scattering
terms have been retained)
\begin{equation}
\frac{\partial n(\nu)}{\partial\tau}=\frac{h}{\me c^2}\frac{1}{\nu^2}
\frac{\partial}{\partial \nu}\nu^4\left\{n^2+\frac{k\te}{\me c^2}
\left[\frac{5}{2}n^2+\frac{42}{5}\nu n\frac{\partial n}
{\partial\nu}+\frac{14}{5}\nu^2 n\frac{\partial^2
n}{\partial\nu^2}-\frac{7}{5}\nu^2\left(\frac{\partial n}{\partial
\nu}\right)^2\right]\right\}.
\label{komp_gen_c}
\end{equation}
It is interesting that only correction terms that are proportional to
$k\te/\me c^2$ appear in this equation. There is no term
proportional to $h\nu/\me c^2$, although such a term is present
in the diffusion equation describing the spontaneous scattering
process, equation (\ref{komp_gen_b}). This is a result of the joint
operation of Compton recoil and Klein-Nishina corrections, which both
contribute to the first two moments of the kernel (eq. [\ref{p_moments}]). 

In the past, many phenomena caused by induced Compton scattering were
investigated in the nonrelativistic approximation, using only the main term in
equation (\ref{komp_gen_c}). One such phenomenon is distortions in the
low-frequency radiation spectra of radio sources \citep{sunyaev71}, which
become large if $kT_b=0.5I_{\nu}\lambda^2\gg \me 
c^2/\tau(1+\tau)$, where $I_{\nu}$ is the intensity of quasi-isotropic
radiation at a wavelength $\lambda$ and $\tau$ is the Thomson
optical depth of the scattering cloud. Particularly interesting is the
case of bright extragalactic radio sources, for which $kT_b\gg \me c^2/\tau$ 
even though $\tau\ll 1$, because $T_b\sim 10^{11}\div 10^{13}$~K. Other
phenomena include plasma heating
\citep{peyraud68,zellev70,levsun71,vinpus72,blasch76,illkom77} and induced 
light-pressure force \citep{levetal72} in the vicinity of 
astrophysical objects emitting low-frequency radiation with high
$T_b$. Obviously, relativistic corrections, described by equation
(\ref{komp_gen_c}), could play an important role in such problems. In
particular, these terms (although small) should play the  
role of viscosity for such phenomena as the formation of shock waves
in the photon spectrum during Bose-condensation of photons \citep{zelsun72}.

Induced Compton scattering may also lead to essentially anisotropic
effects, such as narrowing or spreading (depending on the spectrum of
the radiation) of a radiation beam traversing a plasma
\citep{goletal75,zelsun76}. It may also play a major role in the
interaction of beams of maser radiation having narrow spectra
($\Delta\nu<\Delta\nu_D$, where $\Delta\nu_D=\nu[2(1-\mus)k\te/\me
c^2]^{1/2}$ is the Doppler shift) (Zeldovich et al. 1972). Let us
write (for an infinite homogeneous medium) the integral kinetic
equation that arises in such problems,
\begin{eqnarray}
\frac{\partial n(\nu,\bg{\Omega},\tau)}{\partial\tau} &=&
n(\nu,\bg{\Omega},\tau) \int
n(\nu^\prime,\bg{\Omega^\prime},\tau)\left[\left(\frac{\nu^\prime}{\nu}
\right)^2 K(\nu^\prime,\bg{\Omega^\prime}\rightarrow\nu,\bg{\Omega})-
K(\nu,\bg{\Omega}\rightarrow\nu^\prime,\bg{\Omega^\prime})\right]
d\nu^\prime d\bg{\Omega^\prime}
\nonumber\\
&=& n(\nu,\bg{\Omega},\tau)\int n(\nu^\prime,\bg{\Omega^\prime},\tau) 
K^{\rm ind}(\nu,\bg{\Omega};\nu^\prime,\bg{\Omega^\prime})\,d\nu^\prime
d\bg{\Omega^\prime},
\label{kinetic_ind}
\end{eqnarray}
where $n(\nu)$ is the occupation number in photon phase space, and we
have introduced the new kernel 
$K^{\rm ind}(\nu,\bg{\Omega};\nu^\prime,\bg{\Omega^\prime})$. In the limit
$h\nu\ll m_ec^2$, $h\nu\ll k\te$, which always takes place in
the case of compact sources of low-frequency radiation, a particularly
simple expression for this kernel can be given. Indeed, we can
expand the exponential in the expression (\ref{k}) according to
equation (\ref{q}) of \S 5, and then, taking the second term in the
resulting series out of the exponential, get
\begin{eqnarray}
K^{\rm ind}(\nu,\bg{\Omega};\nu^\prime,\bg{\Omega^\prime})
 &=& \frac{3}{16\pi}\sqrt{\frac{2}{\pi}}\eta^{-3/2}\frac{h\nu^{\prime 2}
(\nu^\prime-\nu)}{\me c^2 g^3}(1-\mus)
\left[1+\mus^2+\left(\frac{1}{8}-\mus-\frac{63}{8}\mus^2+5\mus^3\right)\eta
\right.
\nonumber\\
&&\left.
-\mus(1-\mus^2)\left(\frac{\nu^\prime-\nu}{g}\right)^2
-\frac{3(1+\mus^2)}{8\eta}\left(\frac{\nu^\prime-\nu}{g}\right)^4\right]
\exp{\left[-\frac{(\nu^\prime-\nu)^2}{2g^2\eta}\right]},
\nonumber\\
g &=& |\nu\bg{\Omega}-\nu^\prime\bg{\Omega^\prime}|
=(\nu^2-2\nu\nu^\prime\mus+\nu^{\prime 2})^{1/2}.
\label{k_ind}
\end{eqnarray}

Equation (\ref{k_ind}), like the formula in equation (\ref{k_set}) for the
$K(\nu,\bg{\Omega}\rightarrow\nu^\prime,\bg{\Omega^\prime})$ kernel,
can be used when the plasma is mildly relativistic, $k\te\lesssim
0.1\me c^2$. In the case of nonrelativistic electrons, $k\te\ll\me
c^2$, the correction terms in equation (\ref{k_ind}) can be neglected, and
$\nu^\prime$ can be replaced by $\nu$ everywhere except in the
difference $\nu^\prime-\nu$, which results in the following formula
\begin{equation}
K^{\rm ind}_{\rm nr}(\nu,\bg{\Omega};\nu^\prime,\bg{\Omega^\prime})
=\frac{3}{32}(\pi\eta)^{-3/2}\nu^{-1}\frac{h(\nu^\prime-\nu)}{\me c^2}
(1-\mus)^{-1/2}(1+\mus^2)\exp{\left[-\frac{(\nu^\prime-\nu)^2}{4\nu^2(1-\mus)
\eta}
\right]},
\label{k_ind_nr}
\end{equation}
which was earlier obtained by \cite{zeletal72}.

Integrating the kernel (eq. [\ref{k_ind}]) over all scattering angles
gives the  
corresponding kernel for the isotropic problem. This kernel is, however,
more easily deduced from equation (\ref{p_set}):
\begin{equation}
P^{\rm ind}(\nu;\nu^\prime)=\left(\frac{\nu^\prime}{\nu}\right)^2
P(\nu^\prime\rightarrow\nu)-P(\nu\rightarrow\nu^\prime)
=\sqrt{\frac{2}{\pi}}\eta^{-3/2}
\frac{h(\nu^2+\nu^{\prime 2})(\nu^\prime-\nu)}{\me c^2\nu^2(\nu+\nu^\prime)}
(p_0+p_t),
\label{p_ind}
\end{equation}
where $p_0$ and $p_t$ are given by equation (\ref{pp}). 

The kernel given in equation (\ref{p_ind}) is applicable in the limit
$h\nu\ll k\te\lesssim 0.1\me c^2$ and makes possible to obtain the
differential equation (\ref{komp_gen_c}).

\section{DERIVATION OF THE $K(\nu,\bg{\Omega}\rightarrow
\nu^\prime,\bg{\Omega^\prime})$ KERNEL}

Consider a photon of frequency $\nu$ that propagates in the direction
$\bg{\Omega}$ . We calculate the probability (per unit 
dimensionless time, $\tau$) that the photon will be scattered by a
Maxwellian distribution of electrons into a solid-angle interval
$d\bg{\Omega^\prime}$, with the emergent photon frequency falling in an
interval $d\nu^\prime$.

Our primary goal is to derive a formula that would be a good
approximation for situations in which both the electrons and
photons are mildly relativistic, i.e., when $\eta=k\te/\me c^2,
h\nu/\me c^2\sim 0.1$. Therefore, we make an initial assumption
that $\eta, h\nu/\me c^2\ll \me c^2$. The final formula will
contain correction terms of order up to $\eta^{3/2}$, $\eta^{1/2}
h\nu/\me c^2$, and $(h\nu/\me c^2)^2$. The term of order $(h\nu/\me
c^2)^2$ originates directly from the Klein-Nishina formula for the scattering
cross section. We retain this term in our final expression on purpose
(although it causes the formula to be of a slighly better accuracy in
terms of photon energy that in terms of electron temperature), because
this expression without the temperature-correction terms then
becomes the {\sl exact formula} for the case of scattering of photons
of arbitrary (including $h\nu\gg\me c^2$) energy on nonrelativistic
electrons ($\eta\ll 1$).

Let us introduce the following system of reference: ${\bf Ox}$ points
along $\bg{\Omega}$, ${\bf Oy}$ is in the ($\bg{\Omega}$,
$\bg{\Omega^\prime}$) plane, ${\bf Oz}$ is normal to this scattering
plane. There are two basic equations. The first is the
energy-conservation relation \citep[e.g.,][]{pozetal83}:
\begin{equation}
\frac{\nu^\prime}{\nu}=\frac{1-\bg{\Omega\beta}}{1-\bg{\Omega^\prime\beta}
+(h\nu/\gamma mc^2)(1-\cos{\alpha})},
\label{en_cons}
\end{equation}
where $c\bg{\beta}$ is the electron velocity and
$\cos{\alpha}=\bg{\Omega}\bg{\Omega^\prime}$. The second equation
describes the differential cross section for Compton scattering
\citep{jauroh76,beretal82}:
\begin{equation}
\frac{d\sigma}{d\bg{\Omega^\prime}}=\frac{3\sigma_{\rm T}}
{16\pi\gamma^2}\frac{X} 
{(1-\bg{\Omega\beta})^2}\left(\frac{\nu^\prime}{\nu}\right)^2,
\label{klein}
\end{equation}
where 
\begin{equation}
X=2-\frac{2(1-\cos{\alpha})}{\gamma^2(1-\bg{\Omega\beta})
(1-\bg{\Omega^\prime\beta})}+\frac{(1-\cos{\alpha})^2}{\gamma^4(1-
\bg{\Omega\beta})^2(1-\bg{\Omega^\prime\beta})^2}+\frac{\nu^\prime}{\nu}
\left(\frac{h\nu}{\me c^2}\right)^2\frac{(1-\cos{\alpha})^2}
{\gamma^2(1-\bg{\Omega\beta})(1-\bg{\Omega^\prime\beta})}.
\label{x}
\end{equation}

Introducing the components of the electron velocity
($\beta_x$, $\beta_y$, $\beta_z$), we find 
\begin{eqnarray}
\bg{\Omega\beta}=\beta_x,\,\,\,\bg{\Omega^\prime\beta}=\beta_x
\cos{\alpha}+\beta_y\sin{\alpha}.
\label{omegab}
\end{eqnarray}

Equation (\ref{en_cons}) imposes a link between the different $\bg{\beta}$
components for given $\nu^\prime/\nu$ and $\alpha$:
\begin{equation}
\beta_y=\frac{1}{\sin{\alpha}}\left[\beta_x\left(\frac{\nu}{\nu^\prime}
-\cos{\alpha}\right)+1-\frac{\nu}{\nu^\prime}+\frac{h\nu}{\gamma
\me c^2}(1-\cos{\alpha})\right],
\label{beta_y}
\end{equation}
with
\begin{equation}
\gamma^2=\frac{1}{1-(\beta_x^2+\beta_y^2+\beta_z^2)}.\
\label{gamma}
\end{equation}

In order to calculate $K(\nu,\bg{\Omega}\rightarrow
\nu^\prime,\bg{\Omega^\prime})$, we ought to carry out the following
integration over electron velocities:
\begin{equation}
K(\nu,\bg{\Omega}\rightarrow\nu^\prime,\bg{\Omega^\prime})\equiv\frac{dP}
{d\tau\,d\Omega^\prime\,d\nu^\prime}=\frac{1}{\sigma_{\rm T}}\int
\frac{d\sigma}{d\bg{\Omega^\prime}} (1-\bg{\Omega\beta})
f_{\beta}(\beta_x,\beta_y,\beta_z)\left|\frac{\partial \beta_y}{\partial
\nu^\prime}\right|\,d\beta_x d\beta_z.
\label{k_gen}
\end{equation} 
Here, the factor $(1-\bg{\Omega\beta})$ allows for the
relative velocity of the photon and electron along the direction of
the former's motion \citep{lanlif75},
$f_{\beta}(\beta_x,\beta_y,\beta_z)$ is the electron velocity distribution
function. In equation (\ref{k_gen}), one of the velocity components (say,
$\beta_y$) must be expressed through the other two. The interval,
$d\nu^\prime$, of the photon frequency after scattering can then be related
with the corresponding interval of $\beta_y$; hence the appearence of the
factor $|\partial \beta_y/\partial \nu^\prime|$ in equation (\ref{k_gen}).

For $f_{\beta}$ we substitute the relativistic Maxwellian distribution
function,
\begin{equation}
f_{\beta}=(2\pi\eta)^{-3/2}\left(1+\frac{15}{8}\eta\right)^{-1}
\left(1+\frac{5}{2}\beta^2-\frac{3}{8}\frac{\beta^4}{\eta}\right)
\exp{\left(-\frac{\beta^2}{2\eta}\right)},
\label{fm}
\end{equation}
Here we have retained only correction terms of order $\eta$.

We can proceed with the integration in equation (\ref{k_gen}) on expanding the
integrand in powers of $\beta$. The correct account of temperature
terms of order $\beta^2$ and $\beta^3$ necessitates inclusion of the
corresponding terms in equation (\ref{fm}) for the velocity
distribution (contrary to the statement made in BR70). We also note
that the terms of order $\beta^3$, $\beta h\nu/\me c^2$, and
$(h\nu/\me c^2)^2$, which we are keep throughout, were
neglected altogether in the derivation of BR70.

As follows from equation (\ref{beta_y}), the derivative $\partial
\beta_y/\partial \nu^\prime$ to the first order is
\begin{equation}
\frac{\partial\beta_y}{\partial\nu^\prime}=\frac{\nu}{\nu^{\prime 2}
\sin{\alpha}}(1-\beta_x).
\label{deriv1}
\end{equation}

The last bracketed term in the expression (\ref{beta_y}) gives rise to
additional terms of order $\beta h\nu/\me c^2$ due to the presence of
the factor $1/\gamma$. Using equation (\ref{gamma}) we find
\begin{equation}
\frac{\partial(1/\gamma)}{\partial\nu^\prime}\approx
-\beta_y\frac{\partial\beta_y}{\partial\nu^\prime}\approx
-\beta_y\frac{\nu}{\nu^{\prime 2}\sin{\alpha}},
\label{deriv2}
\end{equation}
which finally yields
\begin{equation}
\frac{\partial\beta_y}{\partial\nu^\prime}=\frac{\nu}{\nu^{\prime 2}
\sin{\alpha}}\left\{1-\beta_x-\frac{h\nu}{\me c^2}\frac{1-\cos{\alpha}}
{\sin^2{\alpha}}\left[\beta_x\left(\frac{\nu}
{\nu^\prime}-\cos{\alpha}\right)+1-\frac{\nu}{\nu^\prime}+\frac{h\nu}{\me c^2}
(1-\cos{\alpha})\right]
\right\}.
\label{deriv3}
\end{equation}

Note that the main Klein-Nishina correction to the scattering
cross section, which is of the order of $h\nu/\me c^2$, is contained in the
factor $(\nu^\prime/\nu)^2$ in equation (\ref{klein}), rather than in the
$X$ function (eq. [\ref{x}]), which describes Doppler aberration and the
Klein-Nishina correction of order $(h\nu/\me c^2)^2$. The reciprocal factor,
$\nu/\nu^{\prime 2}$, enters equation (\ref{deriv3}), so upon
multiplication of $d\sigma/d\bg{\Omega^\prime}$ and $|\partial \beta_y/\partial
\nu^\prime|$ in the integrand of equation (\ref{k_gen}), the presence of the
$O(h\nu/\me c^2)$ correction in
$K(\nu,\bg{\Omega}\rightarrow\nu^\prime,\bg{\Omega^\prime})$ is not explicit. 

Proceeding further with expansions, we obtain
\begin{eqnarray}
\frac{1}{\sigma_{\rm T}}\frac{d\sigma}{d\bg{\Omega^\prime}}
(1-\bg{\Omega\beta})\left|\frac{\partial \beta_y}{\partial 
\nu^\prime}\right|=\frac{3}{16\pi}\frac{1}{\nu\sin{\alpha}}
\left\{1+\mus^2+2s(-\mus+\mus^2)+2\beta_x(-\mus+\mus^2)
\right.
\nonumber\\
\left.
+s^2(1-4\mus+3\mus^2)+2s\beta_x(1-3\mus+2\mus^2)
+\beta_x^2(1-4\mus+3\mus^2)+\beta^2(-1+2\mus-3\mus^2)
\right.
\nonumber\\
\left.
+2s^3(1-3\mus+2\mus^2)+2s^2\beta_x(2-5\mus+3\mus^2)
+2s\beta_x^2(2-5\mus+3\mus^2)+2\beta_x^3(1-3\mus+2\mus^2)
\right.
\nonumber\\
\left.
+2\beta^2 s(-1+4\mus-3\mus^2)+2\beta^2 \beta_x(-1+4\mus-3\mus^2)
+\frac{h\nu}{\me c^2}\left(1-\frac{\nu}{\nu^\prime}+\frac{h\nu}{\me c^2}
(1-\mus)\right)\frac{(1+\mus^2)}{1+\mus}
\right.
\nonumber\\
\left.
-\frac{h\nu}{\me c^2}\beta_x\frac{(1-\mus)(1+\mus^2)}{1+\mus}
+\frac{\nu^\prime}{\nu}\left(\frac{h\nu}{\me c^2}\right)^2(1-\mus)^2\right\},
\label{ind}
\end{eqnarray}
where $\mus=\cos{\alpha}$, 
\begin{equation}
s=\bg{\Omega^\prime\beta}=
\beta_x\frac{\nu}{\nu^\prime}+1-\frac{\nu}{\nu^\prime}+\frac{h\nu}{\me c^2}
(1-\mus),
\label{s}
\end{equation}
\begin{equation}
\beta^2\approx\tilde{\beta^2}=
\beta_x^2+\beta_z^2+\frac{1}{1-\mus^2}\left[\beta_x\left(\frac{\nu}
{\nu^\prime}-\mus\right)+1-\frac{\nu}{\nu^\prime}+\frac{h\nu}{\me c^2}
(1-\mus)\right]^2.
\label{beta2}
\end{equation}
Here $\tilde{\beta^2}$ must be substituted for $\beta^2$ in equation
(\ref{ind}) and in the terms preceeding the exponential in equation
(\ref{fm}).

Note that in the situation of scattering of nonrelativistic electrons
($\eta\ll 1$), equation (\ref{ind}) reduces to a simple formula:
\begin{equation}
\frac{1}{\sigma_{\rm T}}\frac{d\sigma}{d\bg{\Omega^\prime}}
(1-\bg{\Omega\beta})\left|\frac{\partial \beta_y}{\partial 
\nu^\prime}\right|=\frac{3}{16\pi}\frac{1}{\nu\sin{\alpha}}
\left[1+\mus^2+\frac{\nu^\prime}{\nu}\left(\frac{h\nu}{\me c^2}\right)^2
(1-\mus)^2\right],
\label{ind_nr}
\end{equation}
which is valid for arbitrary values of photon energy, including
$h\nu\gg \me c^2$. Our final formula will consequently be exact in the
limit $\eta\ll 1$, as we claimed at the beginning of this section.

Having made this remark, let us return to considering our main situation of
interest, i.e,. with mildly relativistic electrons and
photons. The accuracy of equation (\ref{beta2}) turns out to be insufficent
for describing the exponential factor of the distribution function
(eq. [\ref{fm}]), in which $\beta^2$ is divided by $\eta$. In this factor,
$\beta^2$ need be given accurately to the fifth order, which requires
the inclusion of the factor $1/\gamma$ in equation (\ref{beta_y}) (similarly to
the situation with $\partial\beta_y/\partial\nu^\prime$ above). We
consequently get 
\begin{equation}
\exp\left(-\frac{\beta^2}{2\eta}\right)=\exp\left(-\frac{\tilde{\beta^2}+
\Delta\beta^2}{2\eta}\right)\approx
\left(1-\frac{\Delta\beta^2}{2\eta}\right)\exp\left(-\frac{\tilde{\beta^2}}
{2\eta}\right),
\end{equation}
where
\begin{equation}
\Delta\beta^2=-\frac{1-\cos{\alpha}}{\sin{\alpha}}\frac{h\nu}{\me c^2}\beta_y
\tilde{\beta^2},
\end{equation}
with $\beta_y$ and $\tilde{\beta^2}$ given by equations (\ref{beta_y}) and
(\ref{beta2}), respectively.

Now, having completed all the necessary preparations, we can implement the 
integration in equation (\ref{k_gen}). Integrals of the following type then
appear:
\[
\int \beta_x^k\beta_z^l\exp{\left\{-\frac{1}{2\eta}[\beta_x^2+\beta_z^2+
(a\beta_x+b)^2]\right\}}\,d\beta_x\,d\beta_z,
\]
where $a$ and $b$ are constants set by equation (\ref{beta2}). Such
integrals are readily done (see, e.g., BR70). It is natural to present the
final result in the form of a series in terms of the quantity 
\begin{equation}
\Delta=\nu^\prime-\nu+\nu^\prime\frac{h\nu}{\me c^2}(1-\cos{\alpha}).
\label{delta}
\end{equation}
Indeed, consider the situation with $k\te=0$, when all scattered photons
undergo the same decrement in frequency, owing
to Compton recoil: $\nu/\nu^\prime=1+(1-\cos{\alpha})h\nu/\me c^2$, as 
follows from equation (\ref{en_cons}). This shift corresponds exactly to
$\Delta=0$. If we now allow a nonvanishing electron temperature
($k\te\neq 0$), the scattered profile will be Doppler-broadened near
the recoil-shifted peak of the line. The term $\Delta$ will then measure
the frequency variation relative to this peak and therefore will 
always be of the order of $\nu\eta^{1/2}$, which ensures that the
final expression will be convergent regardless of the proportion of
$h\nu$ and $k\te$ (for comparison, we can consider a similar quantity
$\nu^\prime-\nu$, which is $\sim\nu\eta^{1/2}$ if the Doppler effect is
dominant but $\sim -h\nu^2/\me c^2$ if recoil is more
important). Moreover, note that $\Delta$ enters as an entity in
all the major expressions we obtained above (see
eqs. [\ref{beta_y}], [\ref{deriv3}]--[\ref{beta2}]). After lengthy
calculations, we finally arrive at equation (\ref{k_set}) given in \S 2.

\section{CALCULATION OF THE ANGULAR SCATTERING FUNCTION}

The angular function, $d\sigma/d\mus$, for Compton scattering on
Maxwellian electrons can be directly calculated
using the same formalism that we employed to derive the 
$K(\nu,\bg{\Omega}\rightarrow \nu^\prime,\bg{\Omega^\prime})$ kernel 
in the preceeding section. In fact, the derivation of
$d\sigma/d\mus$ is less tedious than that of
$K(\nu,\bg{\Omega}\rightarrow \nu^\prime,\bg{\Omega^\prime})$. For
this reason, we here calculate the angular function to a better
accuracy than that of our final formula for the kernel
(eq. [\ref{k_set}]). Namely, we are willing to obtain an expression that will
contain correction terms of the order of $(k\te/\me c^2)^2$ and
$(h\nu/\me c^2)(k\te/\me c^2)$ to the Rayleigh angular
function. The term of order $(h\nu/\me c^2)^2$ will also be found,
but this term already follows from equation (\ref{k_nr}) for the
$K_{\rm nr}$ kernel, which, as we recall, is accurate for arbitrary
photon energies in the case of nonrelativistic electrons ($k\te\ll \me c^2$).

In order to find the angular function, the integration over the electron
velocity space needs to be done:
\begin{equation}
\frac{d\sigma}{d\mus}=\frac{2\pi}{\sigma_{\rm T}}\int
\frac{d\sigma}{d\bg{\Omega^\prime}}(1-\bg{\Omega\beta})
f_{\beta}(\beta_x,\beta_y,\beta_z)\,d\beta_x d\beta_y d\beta_z.
\label{ang_gen}
\end{equation} 
Here the differential scattering crosscsection,
$d\sigma/d\bg{\Omega^\prime}$, for given $(\beta_x,\beta_y,\beta_z)$
and $h\nu/\me c^2$ is given by equations (\ref{en_cons}) and
(\ref{klein}), and the relativistic Maxwellian distribution function
is represented by the series given in equation (\ref{fm}). The
principal difference (which simplifies the calculation) of equation
(\ref{ang_gen}) from the similar equation (\ref{k_gen}) is that
$\beta_y$ is now a free parameter, like 
the other components of the electron velocity.

Next, the quantity $(1-\bg{\Omega\beta})d\sigma/d\bg{\Omega^\prime}$
needs to be expanded in powers of $\beta_x$, $\beta_y$, $\beta_z$ (to
fourth order) and $h\nu/\me c^2$ (to second order), as we similarly did
(to a worse accuracy) in the previous section (see eq. [\ref{ind}]). The
resultant expression is rather cumbersome, so we do not give it
here. For the $f_{\beta}$ distribution function, the approximation of
equation (\ref{fm}) is sufficient, because the next order terms,
$O(\eta^2,\eta\beta^2, \beta^4,\beta^6/\eta,\beta^8/\eta^2)$, in the
series $f_{\beta}$ cancel upon integration. 

The integration in equation (\ref{ang_gen}) is connected to the
calculation of standard integrals
\begin{equation}
\int \beta_x^k\beta_y^l\beta_z^m \exp{\left\{-\frac{\beta_x^2+\beta_y^2
+\beta_z^2}{2\eta}\right\}}\,d\beta_x d\beta_y d\beta_z,
\end{equation}
where $k,l$ and $m$ are even numbers [odd terms with respect to one of the
$(\beta_x,\beta_y,\beta_z)$ components vanish upon integration].

The final result is equation (\ref{ang2}).

\section{DERIVATION OF THE $P(\nu\rightarrow\nu^\prime$) KERNEL}

Here we demonstrate how to perform the integral in
equation (\ref{k_int}) with $K(\nu,\bg{\Omega}\rightarrow
\nu^\prime,\bg{\Omega^\prime})$ given by equation (\ref{k_set}) when
$h\nu(h\nu/\me c^2)\ll k\te$.

Write the exponential factor entering equation (\ref{k}) in the form
of a polynomial:

\begin{equation}
\frac{\epsilon^2}{4(1-\mus)\eta}=\frac{(\nu^\prime-\nu)^2}{2g^2\eta}
+\frac{h\nu\nu^\prime}{\me c^2}\frac{(\nu^\prime-\nu)(1-\mus)}{g^2\eta}
+\left(\frac{h\nu\nu^\prime}{\me c^2}\right)^2\frac{(1-\mus)^2}{2g^2\eta}
\label{q}
\end{equation}
where 
\begin{equation}
g^2=2\nu\nu^\prime(1-\mus)+(\nu^\prime-\nu)^2.
\label{g2}
\end{equation}
One can see from equation (\ref{g2}) that $g^2\rightarrow 0$ at
$\nu^\prime=\nu$ when $\mus\rightarrow 1$. The factor $g^2$ enters
the denominator of the first member of the polynomial in equation
(\ref{q}), which 
describes the Doppler broadening. Knowing this property, which means that
$K(\nu,\bg{\Omega}\rightarrow \nu^\prime,\bg{\Omega^\prime})$ is a
$\delta$-function for $\mus=0$, one can immediately make a
prediction that the $P(\nu\rightarrow\nu^\prime)$ kernel resulting
from the integration in equation (\ref{k_int}) over $\mus$ will have a
cusp (a point where a break in the derivative occurs) at 
$\nu^\prime=\nu$. Such a cusp is indeed present in our final
expression for $P(\nu\rightarrow\nu^\prime)$, as we shall see below.

There is no singularity at $\mus=1$ in the second and third members
of the polynomial (\ref{q}), which describe the frequency variation due to
Compton recoil. This suggests that the cusp mentioned above will
remain at the same position, $\nu^\prime=\nu$, regardless of the initial
photon energy ($h\nu$), which indeed proves to be the case \citep[see the
results of Monte Carlo simulations in Fig.~1 of the review by][]{pozetal83}.
 
By assumption ($h\nu\sim k\te$), the main contribution to the
photon frequency increment by scattering comes from the
Doppler effect. This implies that $\nu^\prime-\nu$ is typically $\sim
\nu\eta^{1/2}$. The second and third terms in equation (\ref{q})
are thus infinitesimal of the order of $\eta^{1/2}$ and $\eta$,
respectively. One can therefore take these terms out of the
exponential, which will make possible analytical integration in
equation (\ref{k_int}).

It is convenient to present our final result for $K(\nu,\bg{\Omega}\rightarrow
\nu^\prime,\bg{\Omega^\prime})$ in terms of the quantity
\begin{equation}
\delta=(2\eta)^{-1/2}\frac{\nu^\prime-\nu}{\nu+\nu^\prime}.
\label{delta_t}
\end{equation}
This new variable, describing the relative frequency shift, is similar
to $\epsilon$, in powers of which the original equation (\ref{k_set}) is
written, when $h\nu\rightarrow 0$ (see the main term of eq. [\ref{epsilon2}]
below). Moreover, the combination
$(\nu^\prime-\nu)/(\nu+\nu^\prime)$ arises in a natural way if one 
first calculates the analog of the $P(\nu\rightarrow\nu^\prime)$
kernel for a monoenergetic isotropic electron distribution and then
convolves this quantity with the Maxwellian distribution function (as we do
in \S 6). Indeed, for a given electron speed, $\beta c$, the photon
frequency after scattering can take values in the range
$|\nu^\prime-\nu|/(\nu+\nu^\prime)\le \beta$. From this fact it
becomes immediately clear that $P(\nu\rightarrow\nu^\prime)$ 
must (and indeed does) possess a certain symmetry if expressed in terms of
$\delta$ given by equation (\ref{delta_t}). 

After a few intermediate steps (see eq. [\ref{k_set}]), which are 
\begin{eqnarray}
\frac{\epsilon^2}{\eta} &=& 8\delta^2+4\sqrt{2}(1-\mus)\delta\frac{h\nu}
{\me c^2}\eta^{-1/2}-\frac{16(1+\mus)}{1-\mus}\delta^4\eta+8(1-\mus)\delta^2
\frac{h\nu}{\me c^2}+(1-\mus)^2\left(\frac{h\nu}{\me c^2}\right)^2\eta^{-1} 
\nonumber\\
 &&-8\sqrt{2}(1+\mus)
\delta^3\frac{h\nu}{\me c^2}\eta^{1/2}+2\sqrt{2}(1-\mus)^2\delta
\left(\frac{h\nu}{\me c^2}\right)^2\eta^{-1/2}+...\,,
\label{epsilon2}
\end{eqnarray}
\begin{equation}
\frac{\nu^\prime}{g}=[2(1-\mus)]^{-1/2}\left(1+\sqrt{2}\delta\eta^{1/2}
-\frac{1+\mus}{1-\mus}\delta^2\eta-\frac{\sqrt{2}(1+
\mus)}{1-\mus}\delta^3 
\eta^{3/2}\right)+...\,,
\end{equation}
\begin{mathletters}
we finally obtain
\begin{eqnarray}
K(\nu,\bg{\Omega}\rightarrow\nu^\prime,\bg{\Omega^\prime})=\nu^{-1}\frac{3}
{32\pi}[\pi(1-\mus)\eta]^{-1/2}\exp{\left(-\frac{2\delta^2}{1-\mus}\right)}
\nonumber\\
\times\left\{\left[1+\sqrt{2}\delta\left(1-\frac{h\nu}{k\te}\right)\eta^{1/2}
-4\delta^2\frac{h\nu}{\me c^2}+2\sqrt{2}\delta^3\left(-2+\frac{1}{3}
\left(\frac{h\nu}{k\te}\right)^2\right)\frac{h\nu}{\me c^2}
\eta^{1/2}\right]k_0 
\right.
\nonumber\\
\left.
+\left[1+\sqrt{2}\delta\left(1-\frac{h\nu}{k\te}\right)\eta^{1/2}\right]k_t
+\left[1+\sqrt{2}\delta\left(3-\frac{h\nu}{k\te}\right)\eta^{1/2}\right]k_r
\right\},
\label{k_exp}
\end{eqnarray}
where
\begin{eqnarray}
k_0 &=& 1+\mus^2,
\nonumber\\
k_t &=& \left[\frac{1}{8}-\mus-\frac{63}{8}\mus^2+5\mus^3+\frac
{-1-5\mus-\mus^2+3\mus^3}{1-\mus}\delta^2
+\frac{2(-1+2\mus-\mus^2+2\mus^3)}{(1-\mus)^2}\delta^4\right]\eta,
\nonumber\\
k_r &=& (1+\mus^2)\left(-\frac{1-\mus}{4}+\delta^2\right)
\left(\frac {h\nu}{\me c^2}\right)^2\eta^{-1}.
\end{eqnarray}\
\label{k_exp_set}
\end{mathletters}

Note that we have omitted in equation (\ref{k_exp_set}) the second-order,
$O((h\nu/\me c^2)^2)$, Klein-Nishina correction term, which was present
in the original expression in equation (\ref{k_set}). This is because
in the limit 
we are currently working with ($h\nu\sim k\te$), inclusion of this term
would be inconsistent with the absence of (unknown) terms of the same order,
such as $O(\eta^2)$ or $O((h\nu/\me c^2)^3\eta^{-1})$, in the series
given in equation (\ref{k_exp}). 

The term $K(\nu,\bg{\Omega}\rightarrow \nu^\prime,\bg{\Omega^\prime})$ in the
form (\ref{k_exp_set}) is easily integrated over the 
scattering angle using the change of variables ($\mus\rightarrow t$) 
$2\delta^2/(1-\mus)=t^2$, which results in the final equation
(\ref{p_set}) for the $P(\nu\rightarrow\nu^\prime)$ kernel for the isotropic
problem.

\section{DIRECT CALCULATION OF THE $P(\nu\rightarrow\nu^\prime$)
KERNEL IN THE CASE OF $h\nu\ll k\te$}

Given is an isotropic field of electromagnetic radiation of frequency
$\tilde{\nu}$. Its spectrum (number of photons per unit solid angle,
unit frequency interval, and unit detector area) is
\begin{equation}
\frac{d N(\nu)}{d\Omega\,d\nu}=\frac{\delta(\nu-\tilde{\nu})}{4\pi}.
\end{equation}

Consider scattering of the radiation on an electron moving at a
speed of $v=\beta c$. The energy of the photons is assumed to be low enough 
($h\tilde{\nu}\ll\me v^2$) that Compton recoil can be ignored. We
consider electrons that are not too relativistic:
$(h\tilde{\nu}/\me c^2)\gamma\ll 1$, where $\gamma=(1-\beta^2)^{-1/2}$;
thus Klein-Nishina relativistic corrections are not significant. For the
moment, we ignore induced scattering. Consider the situation in the electron
rest frame. In this frame, the spectral intensity of the incident radiation
is direction dependent: 
\begin{equation}
\frac{d N_0(\mu_0,\nu_0)}{d\Omega_0\,d\nu_0}=\left(\frac{\nu_0}{\nu}\right)^2
\frac{d N(\nu)}{d\Omega\,d\nu},
\end{equation}
where $\mu$ is the cosine of the angle between the velocity of the
electron and the direction of propagation of the photon. Quantities that
are measured in the electron rest frame (with subscript ``0'')  are
related to the corresponding quantities measured in the laboratory frame via
the Lorentz transformations: 
\begin{equation}
\mu_0=\frac{\mu-\beta}{1-\beta\mu},\,\,\nu_0=\frac{\nu}{\gamma(1+\beta\mu_0)}.
\label{lorentz}
\end{equation}
This leads to 
\begin{equation}
\frac{dN_0(\mu_0,\nu_0)}{d\Omega_0\,d\nu_0}=\frac{1}{4\pi\gamma^3(1+\beta
\mu_0)^3}\delta\left(\nu_0-\frac{\tilde{\nu}}{\gamma(1+\beta\mu_0)}\right).
\end{equation}

Under the given assumptions, the scattering can be treated in the
Thomson limit (the photon frequency does not change) in the electron
rest frame. Therefore, the number of photons scattered into an
interval $d\Omega_0$ of solid angle in a unit time is 
\begin{eqnarray}
\frac{d N_0(\mu_0,\nu_0)}{dt\,d\Omega_0\,d\nu_0}=\frac{3\sigma_{\rm T} c}{16}
\int_{-1}^{1}d\mu_0^\prime (3+3\mu_0^2\mu_0^{\prime 2}-\mu_0^2-\mu_0^{\prime
2})\frac{d N(\mu_0^\prime,\nu_0)}{d\Omega_0^\prime\,d\nu_0}
\nonumber\\
=\frac{3\sigma_{\rm T}\nu_0}{64\pi\gamma\beta\tilde{\nu}^2}\left[3+
\frac{3\mu_0^2}{\beta^2}\left(1-\frac{\tilde{\nu}}{\gamma\nu_0}\right)^2
-\mu_0^2-\frac{1}{\beta^2}\left(1-\frac{\tilde{\nu}}{\gamma\nu_0}\right)^2
\right].
\end{eqnarray}

The reverse transition to the laboratory frame can be performed using
equation (\ref{lorentz}) by the formula 
\begin{eqnarray}
\frac{d N(\mu,\nu)}
{dt\,d\Omega\,d\nu}=\frac{dt_0}{dt}\frac{d\mu_0}
{d\mu}\frac{d\nu_0}{d\nu}\frac{d N_0(\mu_0,\nu_0)}{d t_0\,d\Omega_0\,d\nu_0}=
\frac{1}{\gamma^2(1-\beta\mu)}\frac{d N_0(\mu_0,\nu_0)}
{dt_0\,d\Omega_0\,d\nu_0}~~~~~~~~~~~~~~~~~~~~~~~~~~~~~~~~~~~~~~~~~~~~
\nonumber\\
=\frac{3\sigma_{\rm T} cu}{64\tilde{\nu}\pi\gamma^2\beta}
\left[3+\frac{3}{\beta^2}\left(\frac{\mu-\beta}{1-\beta\mu}\right)^2
\left(1-\frac{1}{\gamma^2(1-\beta\mu)u}\right)^2-
\left(\frac{\mu-\beta}{1-\beta\mu}\right)^2-\frac{1}{\beta^2}
\left(1-\frac{1}{\gamma^2(1-\beta\mu)u}\right)^2\right],
\label{p_beta1}
\end{eqnarray}
where $u=\nu/\tilde{\nu}$.

The derivation given above is analogous to the one in
\citep{sazsun98}. The difference is that that paper considered the 
scattering of a Planckian spectrum\footnote{A formula similar to
equation (\ref{p_beta1}) made it possible to find relativistic
corrections to the 
amplitude of CMB distortions in the direction of clusters of galaxies.
These corrections are of the order of $(k\te/\me c^2)^2$,
$(V/c)^2$, and $(V_r/c)\times (k\te/\me c^2)$, where $V$
and $V_r$ are the cluster peculiar velocity and its component along
the line of sight, respectively.}.

For a given $\mu$, the photon frequency change, $u$, ranges in 
$[(1-\beta)/(1-\beta\mu), (1+\beta)/(1-\beta\mu)]$. Integrating
equation (\ref{p_beta1}) over the solid angle of emergence of the
scattered photon yields the spectrum that forms as a result of scattering of
monochromatic radiation on electrons moving at a speed of $\beta$. For
a given $u$, the integration limits are 
\begin{equation}
\left\{
 \begin{array}{rcll}
  -1 & \le \mu \le & (u-1+\beta)/(\beta u),\,\,\, & u<1; \\
  (u-1-\beta)/(\beta u) & \le \mu \le & 1,\,\,\,  & u>1.\\
 \end{array}
\right.
\end{equation}

The result is 
\begin{eqnarray}
\left(\frac{dN(\nu)}{dt\,d\nu}\right)_{+,-}=\frac{3\sigma_{\rm T}c}
{32\tilde{\nu}u\gamma^2\beta^6}\left\{\mp(u-1)\left(\frac{u^2+6u+1}
{\gamma^4}+4u\right)
\right.
\nonumber\\
\left.
+2u(u+1)\left[2\beta\left(\frac{3}{\gamma^2}+\beta^4\right)
+\frac{3-\beta^2}{\gamma^2}
\left(\ln{\frac{1-\beta}{1+\beta}}\pm \ln{u}\right)\right]\right\},
\label{p_beta2}
\end{eqnarray}
where the subscript plus sign corresponds to the values $u>1$, and
the minus sign to $u<1$. Equation (\ref{p_beta2}) was earlier derived by
\cite{faretal97}. The derivation of these authors differs from the one above
in the order of integrations: they first considered the scattering of a
beam of photons by a beam of electrons with a given angle between the two,
and then implemented the integration over this angle.

Let us now make a transition from a single electron to an ensemble of
electrons with density $N_{\rm e}$ and introduce a quantity,
related to the result of equation (\ref{p_beta2}), that gives the 
probability of a scattering event calculated per unit dimensionless
time $\tau=\sigma_{\rm T} N_{\rm e} ct$, defined as
\begin{equation}
P(\tilde{\nu}\rightarrow\nu,\beta)=\frac{1}{\sigma_{\rm T} c}\frac{d N(\nu)}
{dt\,d\nu}.
\label{p_u_beta}
\end{equation}

Equation (\ref{p_u_beta}) describes the kernel of the integral
kinetic equation for the problem of Comptonization on
monoenergetic electrons. Let us mention its basic properties: (1)
$P(\nu\rightarrow\tilde{\nu},\beta)=(\tilde{\nu}/\nu)^2
P(\tilde{\nu}\rightarrow\nu,\beta) $, (2)
$\int_{\nu_{min}}^{\nu_{max}}P\,d\nu=1$, and (3) 
$\int_{\nu_{min}}^{\nu_{max}}(\tilde{\nu}/\nu-1)P\,d\nu=4(\gamma^2-1)/3$, where
$\nu_{min}=(1-\beta)/(1+\beta)$, $\nu_{max}=(1+\beta)/(1-\beta)$ are
the minimal and maximal possible values of the photon frequency after  
scattering. Property 1 ensures that the detailed balance principle is
obeyed (see also eq. [\ref{p_balance}] for the case of Maxwellian
electrons). In property 2, photon conservation manifests itself. Property
3 is a well-known relation (see, e.g., eq. [2.33] in the review by
Pozdnyakov et 
al. 1983), which describes the rate at which the electrons transfer
energy to the photons. There exists a more general expression than
equation (\ref{p_beta2}) (which was obtained in the Thomson limit) for the
$P(\tilde{\nu}\rightarrow\nu,\beta)$ kernel (see eqs. [8.1.8] and
[8.1.12] in Nagirner \& Poutanen 1994, and references therein), which
is valid for arbitrary photon and electron energies.

In Figure~17, we present examples of spectra described by equations
(\ref{p_beta2}) and (\ref{p_u_beta}). Their characteristic feature is
the presence of a cusp at $\nu/\tilde{\nu}=1$. Note that the right
wing of the line contains more photons ($\approx 1/2+69\beta/140$)
than the left one ($\approx 1/2-69\beta/140$), the asymmetry becoming
more pronounced as the electrons get more relativistic. Such spectra
will build up when isotropic monochromatic radiation is scattered
off an optically thin cloud of electrons that are moving isotropically
at the same speed. Note that equation (\ref{p_beta2}) retains valid for
ultrarelativistic ($\gamma\gg 1$) electrons \citep[the case that
interested][]{faretal97}.

\begin{figure*}[tb]
\epsfxsize=16.5cm
\epsffile[-110 190 685 700]{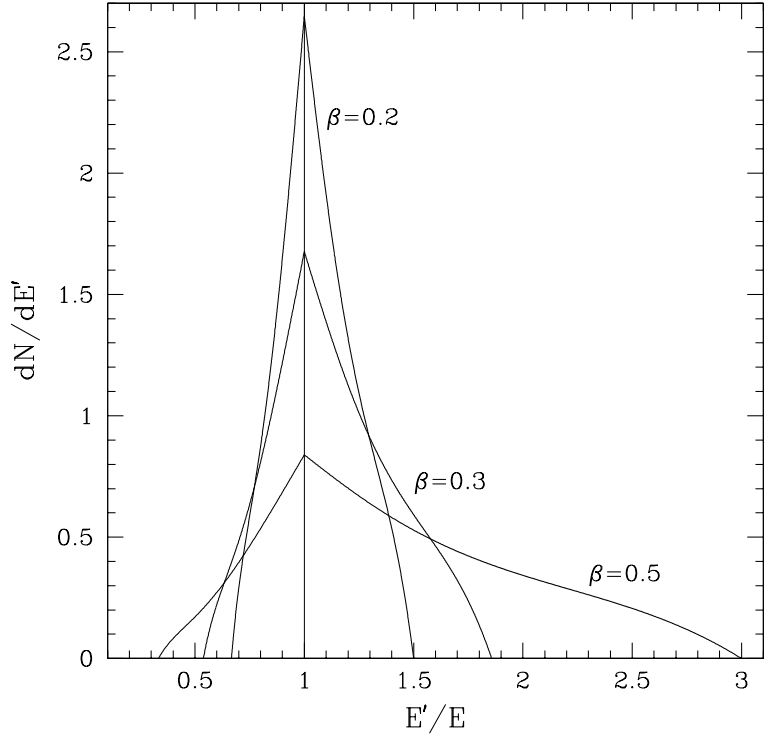}
\caption{
Spectra resulting from the single scattering of isotropic
monochromatic radiation on an ensemble of electrons moving isotropically at
a given speed, as calculated from formula (\ref{p_beta2}) for different
values of the electron velocity ($v=\beta c$). Compton recoil is ignored,
which corresponds to the limit $h\nu\ll \me v^2$.
}
\end{figure*}

In the nonrelativistic limit ($\beta \ll 1$), the frequency is changed upon 
scattering by a small amount: $|u-1|/(u+1) \ll 1$. This allows us, by 
expanding equation (\ref{p_beta2}) in powers of $\beta$ and 
\begin{equation}
\xi=\frac{u-1}{u+1},
\label{Delta}
\end{equation}
to derive the formula
\begin{eqnarray}
P(\tilde{\nu}\rightarrow\nu,\beta)=\frac{1}{\tilde{\nu}\beta}\left[\left(
\frac{11}{20}-\frac{73}{140}\beta^2+O(\beta^4)\right)+\frac{|\xi|}{\beta}
\left(-\frac{3}{4}+\frac{3}{4}\beta^2+O(\beta^4)\right)
\right.
\nonumber\\
\left.
+\left(\frac{\xi}{\beta}\right)^2\left(\frac{11}{20}\beta^2+O(\beta^4)\right)
+\left(\frac{|\xi|}{\beta}\right)^3
\left(\frac{1}{2}-\frac{7}{4}\beta^2+O(\beta^4)\right)
+\left(\frac{|\xi|}{\beta}\right)^5\left(-\frac{3}{10}+\frac{17}{10}\beta^2
+O(\beta^4)\right)
\right.
\nonumber\\
\left.
+\left(\frac{|\xi|}{\beta}\right)^7\left(-\frac{51}{70}\beta^2+O(\beta^4)
\right)\right](1+\xi),
\label{p_beta3}
\end{eqnarray}
which approximates well the scattered photon distribution. The series
given in equation (\ref{p_beta3}) ensures photon conservation to an
accuracy of $\int P\,d\nu=1+O(\beta^4)$, or, putting the same
expression in explicit form, $\int P\,d\nu=1+19\beta^4/300+...\,$.

The spectrum that forms as a result of the single scattering of isotropic
monochromatic radiation on a group of electrons with a given isotropic
distribution of velocities, $f_{\beta}$, is described in general by the
formula 
\begin{equation}
P(\tilde{\nu}\rightarrow\nu)=\int P(\tilde{\nu}\rightarrow\nu,\beta)
f_{\beta}\,\beta^2\,d\beta, 
\label{p_f}
\end{equation}
where $P(\tilde{\nu}\rightarrow\nu,\beta)$ is governed by equations
(\ref{p_beta2}) and (\ref{p_u_beta}). In the case of a thermal plasma, one
must insert the relativistic Maxwellian distribution function into equation
(\ref{p_f}). The result can then be evaluated numerically.

In the present study, we are interested in the mildly relativistic case,
$\eta=k\te/\me c^2\lesssim 0.1$. In this limit, we can derive
$P(\tilde{\nu}\rightarrow\nu)$ analytically by making use of the 
approximate equation (\ref{p_beta3}). To this end, the $f_{\beta}$
function must be expanded in terms of $\beta$. In order to obtain 
the result with the accuracy we need, it is enough to retain only
relativistic correction terms of order $\eta$ in this series, i.e., to
substitute equation (\ref{fm}) of \S 3 for $f_{\beta}$.

The integral in equation (\ref{p_f}) is to be performed in the range of
values $\beta\ge \beta_m=|\xi|=|(u-1)/(u+1)|$. Upon elementary
calculations, we arrive at equation (\ref{pt}) (with the transition
from $\tilde{\nu}$, $\nu$ to $\nu$, $\nu^\prime$), which was our aim.

\begin{acknowledgements}
This work was supported in part by the Russian Foundation for Basic
Research through grants 97-02-16264, 00-15-96649, and 00-02-16681.
\end{acknowledgements}

\end{document}